\documentclass[aps,prd,twocolumn,superscriptaddress,nofootinbib]{revtex4-2}

\usepackage[utf8]{inputenc}
\usepackage[T1]{fontenc}
\usepackage{amsmath}
\usepackage{amssymb}
\usepackage{amsfonts}
\usepackage{bm}
\usepackage{graphicx}
\usepackage{float}
\usepackage{xspace}
\usepackage{placeins}

\begin{document}

\title{Effective kinematic variables for searching for dark matter mediators in single top quark production processes}

\author{E.~E.~Boos}
\email{boos@theory.sinp.msu.ru}
\affiliation{Skobeltsyn Institute of Nuclear Physics, M.V.~Lomonosov Moscow State University, 119991 Moscow, Russia}

\author{V.~E.~Bunichev}
\email{bunichev@theory.sinp.msu.ru}
\affiliation{Skobeltsyn Institute of Nuclear Physics, M.V.~Lomonosov Moscow State University, 119991 Moscow, Russia}

\author{L.~V.~Dudko}
\email{dudko@sinp.msu.ru}
\affiliation{Skobeltsyn Institute of Nuclear Physics, M.V.~Lomonosov Moscow State University, 119991 Moscow, Russia}

\date{\today}

\begin{abstract}
Effective kinematic variables are proposed in the rest frame of the common cluster of the top quark and mediator decay products for searching for and determining the type of dark matter mediator particles and measuring their masses in single top quark production processes. The efficiency of the variables is confirmed by full Monte Carlo simulation taking into account the response of the LHC detectors.
\end{abstract}

\maketitle

\section{Introduction}
\label{intro}

\begin{figure}[H]
	\centering
	\includegraphics[width=0.3\textwidth]{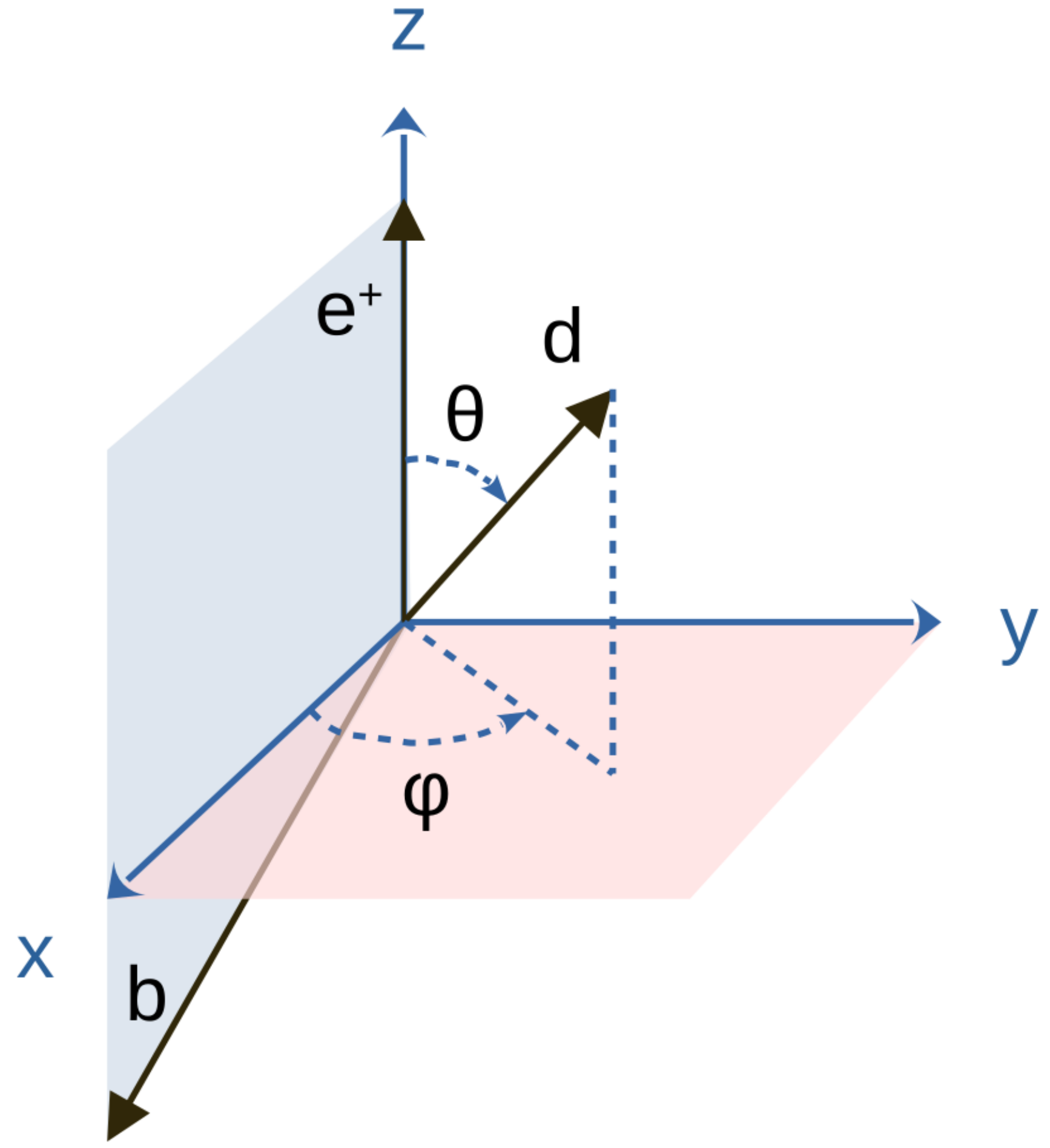}
	\caption{Quantization axis of the top quark spin ($d$-quark direction) in the rest frame of the $t$-quark for the $t$-channel single production process within the SM.}
	\label{fig:coord_param1}
\end{figure}

One of the central problems of modern high-energy physics is the search for dark matter (DM) particles. Astrophysical observations indicate that the vast majority of the Universe is filled with a special type of matter---dark matter. A promising method of investigation is the search for signals from dark matter in accelerator experiments. At the Large Hadron Collider (LHC), processes with large missing transverse energy, which is carried away by dark matter particles, are analyzed to detect processes of dark matter particle production. In popular scenarios, it is assumed that the interaction of the SM fields with the dark matter fields is mediated by intermediate particles (mediators)~\cite{Abdallah:2015ter,Albert:2016osu,Albert:2017onk,Alexander:2016aln,Boveia:2016mrp,Buchmueller:2013dya,Buchmueller:2014yoa,Buckley:2014fba,Chala:2015ama,Choudhury:2015lha,DeSimone:2016fbz,Fairbairn:2014aqa,Haisch:2015ioa,Harris:2014hga,Harris:2015kda,Kahlhoefer:2017dnp,Lebedev:2014bba,Malik:2014ggr,Xiang:2015lfa,Beacham:2019nyx,Abercrombie:2015wmb,Abasov:2024nec}. The cases of scalar, vector, and tensor mediators are considered. The interaction parameters of scalar mediator particles with SM fermions are proportional to the masses of these fermions. Therefore, the search for mediator particles in processes involving massive third-generation fermions, such as the top quark, is theoretically well-motivated and of particular interest. In addition, in single production processes, the top quark can be produced strongly polarized, which is due to the (V-A) structure of the vertices of such interactions~\cite{Jezabek:1994zv,Jezabek:1994qs}. During the top quark decay, its initial polarization is transferred to the decay products and manifests itself in the energy spectra of the decay particles, as well as in spin correlations between the initial and final states. Within the SM, the positively charged lepton from the top quark decay in its rest frame tends to follow the direction of the top quark spin. We recall that in the $t$-channel process of single top quark production, in its rest frame, the spin direction strongly correlates with the $d$-quark momentum~\cite{Mahlon:1996pn,Mahlon:1999gz} and the charged lepton momentum, proportional to the factor $(1 + \cos\theta)$, as shown in Fig.~\ref{fig:coord_param1}.

The SM analytical expression for the differential cross section of the complete production process with the subsequent top quark decay ($2\to 4$) in its rest frame, as a function of the charged lepton energy and two angles of orientation of the top quark spin quantization axis, has the form~\cite{Boos:2019tim}:
\begin{align}\label{totalcrossecsm}
	&\frac{d\sigma_{SM}(\hat{s})_{u b \to d b \nu e^+}}{d\epsilon\cdot d\cos\theta\cdot d\phi}~=~\nonumber\\ 
	&\frac{\alpha^2\cdot V_{ud}^2\cdot V_{tb}^2}{8\cdot3\cdot\sin^4{\Theta_W}\cdot m_W^2\cdot (1 - r_w^2)^2(1+2r_w^2)}\nonumber\\
	&\quad\times\frac{(\hat{s}-m_t^2)^2}{\hat{s}(\hat{s}-m_t^2+m_W^2)}\cdot (1-\epsilon)\epsilon\cdot(1+\cos\theta),
\end{align}
where $\epsilon = 2E_{e^+}/m_t$, $\epsilon_{max} = 1$, $\epsilon_{min} = r_w^2$, $r_w=m_W/m_t$.

\FloatBarrier

For the parametrization, the coordinate system shown in Fig.~\ref{fig:coord_param1} is used. Here $\theta$ is the angle between the charged lepton momentum and the direction of the top quark spin quantization axis (which coincides with the $d$-quark direction), and $\phi$ is the angle in the plane perpendicular to the lepton momentum, measured from the line of intersection with the plane formed by the top quark decay products.

Any deviation from the corresponding SM distribution profiles in the experimental data will indicate the manifestation of new physics.

\section{Effective Lagrangians of DM models}
\label{sec:Lag_DM}

We have considered several of the most general scenarios involving dark matter particles and their mediators. We have assessed the possibility of identifying mediator particles and determining their properties in single top quark production processes for the cases of scalar, pseudoscalar, and vector mediators. The interaction of mediators with SM fermions and DM particles in the most general case is described by the following effective Lagrangians:

\textbf{Scalar mediator $\phi$:}
\begin{equation}
	\mathcal{L}_{\rm int}^{\phi} = -\xi \sum_i \frac{m_i}{v} \phi \bar{\psi}_i \psi_i - g_D \phi \bar{\chi} \chi,
\end{equation}
where $\xi$ is the interaction parameter of the scalar mediator with SM fermions, $v = 246$~GeV is the vacuum expectation value of the Higgs field, and $g_D$ is the interaction parameter of the mediator with DM particles $\chi$.

\textbf{Pseudoscalar mediator $\tilde{\phi}$:}
\begin{equation}
	\mathcal{L}_{\rm int}^{\tilde{\phi}} = -i\xi \sum_i \frac{m_i}{v} \tilde{\phi} \bar{\psi}_i \gamma_5 \psi_i - i g_D \tilde{\phi} \bar{\chi} \gamma_5 \chi.
\end{equation}

\textbf{Vector mediator $A'_\mu$ (with photon-like interaction):}
\begin{equation}
	\mathcal{L}_{\rm int}^{A'} = -\varepsilon e A'_\mu j^\mu_{\rm EM} - e_D A'_\mu j^\mu_{\rm DM},
\end{equation}
where $\varepsilon e$ is the interaction parameter of the vector mediator with the SM electromagnetic current $j^\mu_{\rm EM}$, and $e_D$ is the interaction parameter with the dark matter current $j^\mu_{\rm DM}$.

Below, we investigate in detail how the contributions of various mediators affect the profiles of kinematic distributions in single top quark production processes. We will consider normalized distributions in order not to fix specific scenarios that impose constraints on the mediator--top-quark interaction parameters.

\section{Choice of reference frame and derivation of differential cross sections}
\label{sec:cluster_cross}

To study spin correlations in processes of associated DM and single top quark production, it is necessary to choose an appropriate reference frame. If we could separate the neutrino momentum from the mediator cluster momentum, it would be possible to reconstruct the top quark rest frame. However, in practice this task is difficult to implement, since neutrinos and DM particles are not registered by the detector and are determined as the total missing four-momentum. Taking this circumstance into account, in Refs.~\cite{Boos:2025ecs,Boos:2026wtn} a method was proposed in which all kinematic distributions are constructed in the rest frame of the common cluster corresponding to the top quark and mediator decay products. The four-momentum of this cluster is constructed from the total missing four-momentum of the system, the four-momentum of the charged lepton, and the four-momentum of the most energetic $b$-quark.

\begin{figure}[H]
	\centering
	\includegraphics[width=0.3\textwidth,clip]{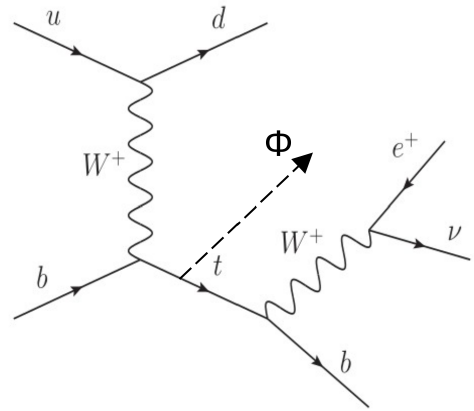}
	\caption{$t$-channel single top quark production process involving a mediator particle.}
	\label{fig3}
\end{figure}

First, we consider in more detail the scenario involving a scalar mediator. In this model, in contrast to the SM process, a scalar particle is emitted from the virtual top quark line (Fig.~\ref{fig3}) with its subsequent decay into a fermion pair of dark matter particles. In the rest frame of the common cluster corresponding to the top quark and mediator decay products, the matrix element of the polarized top quark decay $t \to b \nu e^+$ has the form:
\begin{equation}\label{melement0}
\begin{split}
	|M_{t \to b \nu e^+}|^2 = \frac{g^4}{\left(2(p_{\nu} p_{e^+})-m_W^2\right)^2 + \Gamma_W^2 m_W^2} \\
	\times (p_b p_{\nu}) \cdot \big( (p_{e^+} p_t) - (p_{e^+} s) \cdot m_t \big),
\end{split}
\end{equation}
where $s$ is the top quark spin vector (dimensionless polarization vector). In the common cluster frame it is expressed through the top quark momentum $\mathbf{p}_t$ and the vector of the spin quantization axis direction $\boldsymbol{\eta}$:
\begin{equation}
	s_0 = \frac{(\boldsymbol{\eta} \mathbf{p}_t)}{m_t}, \quad \mathbf{s} = \boldsymbol{\eta} + \frac{\mathbf{p}_t \cdot (\boldsymbol{\eta} \mathbf{p}_t)}{m_t(E_t + m_t)}.
\end{equation}
In this frame the top quark is not at rest, which leads to additional angular dependences.

In the cluster frame corresponding to the top quark and mediator decay products, we introduce the coordinate system described above for the SM case~\cite{Boos:2019tim}. We parametrize the top quark spin direction, as well as the three-dimensional momenta of the top quark decay products, using the angles $\theta$ and $\phi$:
\begin{align}\label{param}
	&\boldsymbol{\eta}=(\sin\theta\cos\phi,\sin\theta\sin\phi,\cos\theta),\nonumber\\
	&\mathbf{p}_{e^+}=|\mathbf{p}_{e^+}|\cdot(0,0,1),\nonumber\\
	&\mathbf{p}_b=|\mathbf{p}_b|\cdot(\sin\theta_{be},0,\cos\theta_{be}),\nonumber\\
	&\mathbf{p}_t\simeq \mathbf{p}_b + 2\mathbf{p}_{e^+},
\end{align}
\begin{align}\label{scalar}
	&(\boldsymbol{\eta}\mathbf{p}_t) = |\mathbf{p}_b|\sin\theta_{be}\sin\theta\cos\phi\nonumber\\
	&\qquad+(|\mathbf{p}_b|\cos\theta_{be}+2|\mathbf{p}_{e^+}|)\cos\theta,
\end{align}
where the angle $\phi$ is expressed through other angular parameters of the system:
\begin{equation}
	\phi = \arccos\left( \frac{\cos\theta_{bd} - \cos\theta_{be} \cos\theta}{\sin\theta_{be} \sin\theta} \right).
\end{equation}
Substituting these expressions into the matrix element and performing algebraic transformations, for the scenario with a scalar mediator we obtain:
\begin{equation}\label{eq:ME_scalar}
	|M_{t \to b \nu e^+}|^2 = F_1 \cdot (1 + \cos\theta) + F_2 \cdot (F_3 - \sin\theta \cos\phi),
\end{equation}
where $F_1$, $F_2$, $F_3$ are functions of the positron and $b$-quark momenta, as well as the angle between them. For a qualitative estimate, one can use approximate expressions in which the positron and neutrino momenta are assumed to be equal:
\begin{equation}
	F_1 \simeq \frac{m_t^2}{4} \cdot r, \quad F_2 \simeq \frac{m_t^2}{4} \cdot \sqrt{1-r^2}, \quad F_3 \simeq \sqrt{\frac{1-r}{1+r}},
\end{equation}
where $r = m_t/(4E_{e^+})$.

The first term preserves the correlation $(1+\cos\theta)$ characteristic of the SM, while the second term, proportional to $\sin\theta\cos\phi$, is a direct consequence of mediator emission. Upon integration over $\cos\theta$, the dependence on $\phi$ takes the form ($-\cos\phi$), increasing with growing $\phi$.

Now we consider the scenario with a pseudoscalar mediator. As in the previous scenario, here in the $t$-channel process a virtual top quark is produced, which emits a mediator and decays into a positron, neutrino, and $b$-quark. The presence of the $\gamma_5$ operator in the vertex of interaction between top quarks and the pseudoscalar mediator leads to a change in the direction of the virtual top quark momentum relative to the production system to the opposite one compared to the scalar mediator case. Substituting (\ref{param}) into (\ref{melement0}) and taking into account the change in the sign of the top quark momentum, we obtain the expression for the matrix element of the top quark decay in the cluster frame corresponding to the top quark and pseudoscalar mediator decay products:
\begin{equation}\label{eq:ME_pseudo}
	|M_{t \to b \nu e^+}|^2 = F_1 \cdot (1 + \cos\theta) + F_2 \cdot (F_3 + \sin\theta \cos\phi),
\end{equation}
where the sign in front of the $\sin\theta\cos\phi$ term changes to the opposite. Accordingly, after integration over $\cos\theta$, a distribution proportional to ($+\cos\phi$) is obtained, decreasing with growing $\phi$. This fundamental difference allows one to identify the mediator type.

For the vector mediator, the situation becomes more complicated, since emission can occur from various fermion lines, and the analytical form of the matrix element becomes cumbersome. Therefore, the analysis for this case will be carried out at the level of numerical simulation.

\section{Mutual orientation of the production and decay systems of the top quark}
\label{sec:DM_geometry}

\begin{figure}[H]
	\centering
	\includegraphics[width=0.4\textwidth]{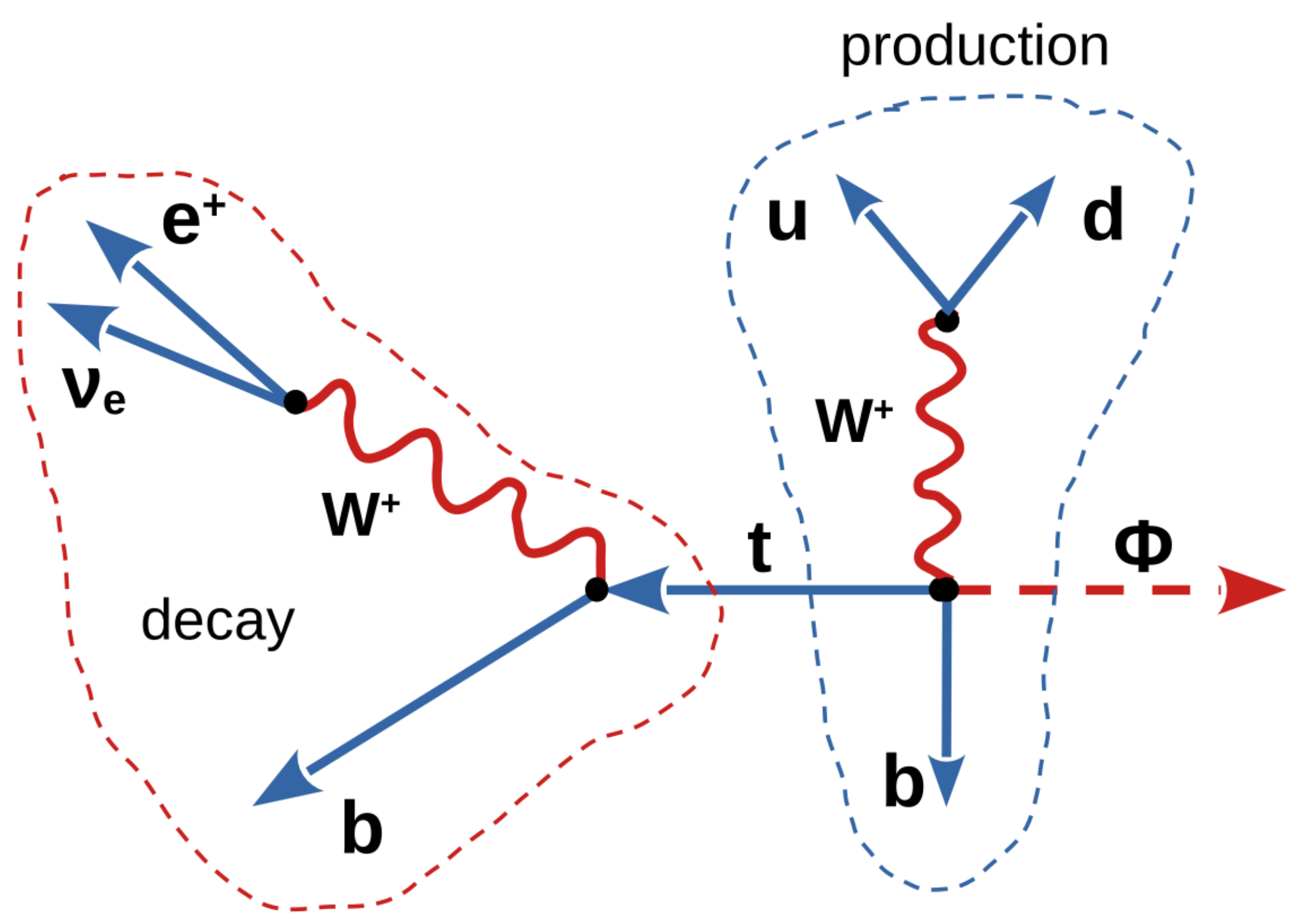}
	\caption{Mutual orientation of the production and decay systems of the top quark in the cluster frame corresponding to the top quark and scalar mediator decay products.}
	\label{fig:DM_orientation_scalar}
\end{figure}

\begin{figure}[H]
	\centering
	\includegraphics[width=0.3\textwidth]{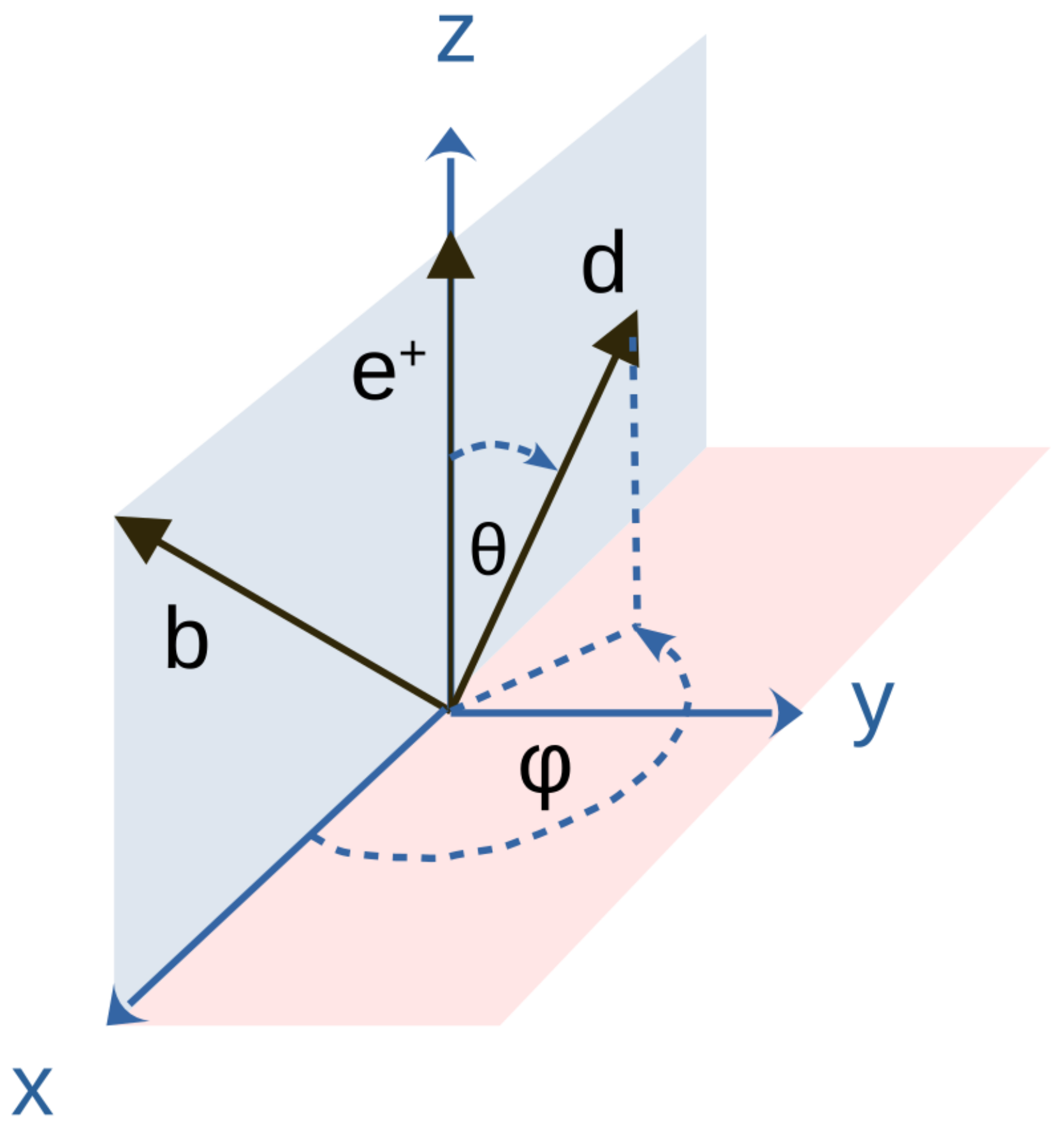}
	\caption{Quantization axis of the top quark spin ($d$-quark momentum direction) in the cluster frame corresponding to the top quark and scalar mediator decay products.}
	\label{fig:DM_spin_scalar}
\end{figure}

\begin{figure}[H]
	\centering
	\includegraphics[width=0.4\textwidth]{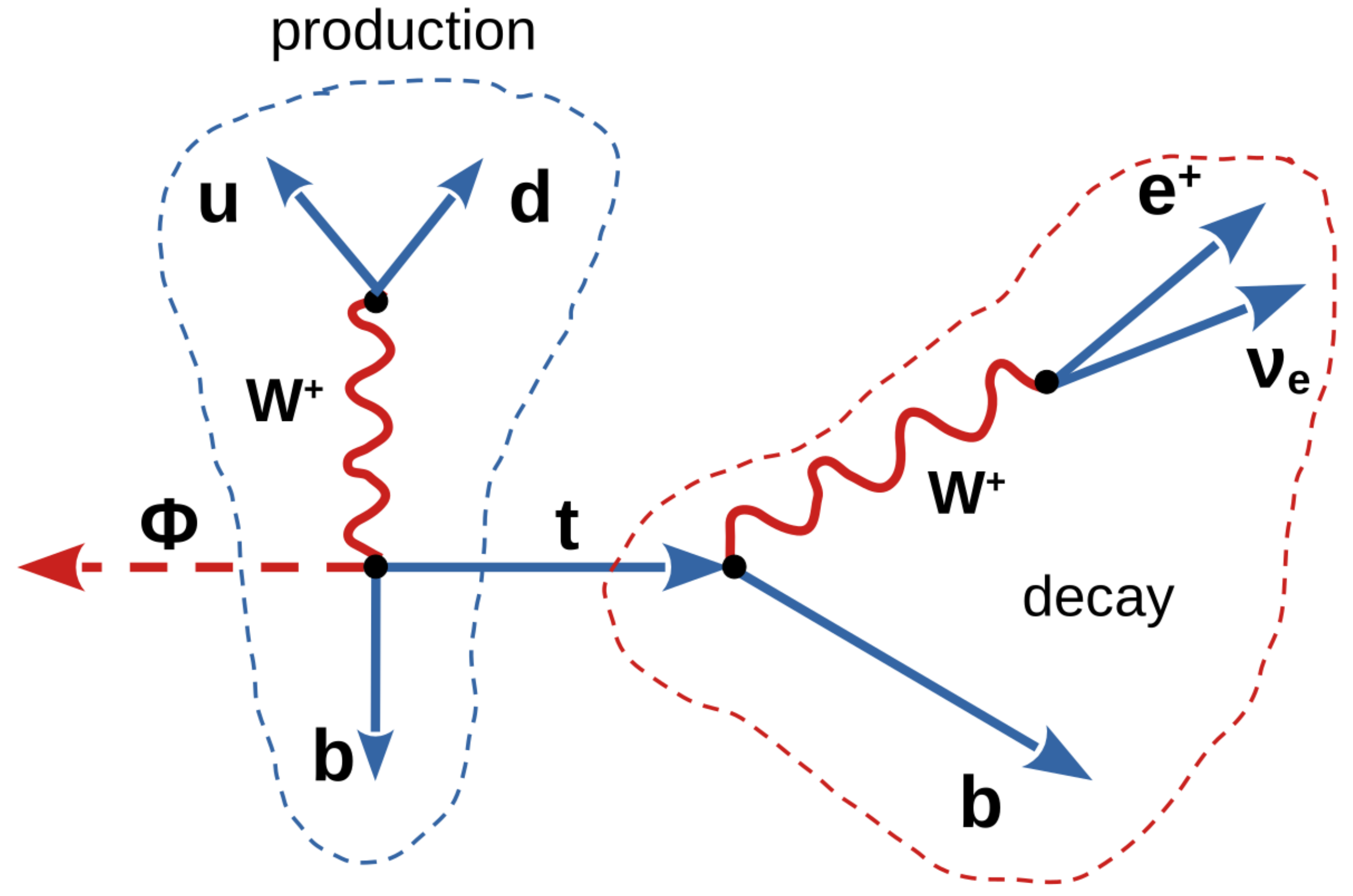}
	\caption{Mutual orientation of the production and decay systems of the top quark in the cluster frame corresponding to the top quark and pseudoscalar mediator decay products.}
	\label{fig:DM_orientation_pseudo}
\end{figure}

\begin{figure}[H]
	\centering
	\includegraphics[width=0.3\textwidth]{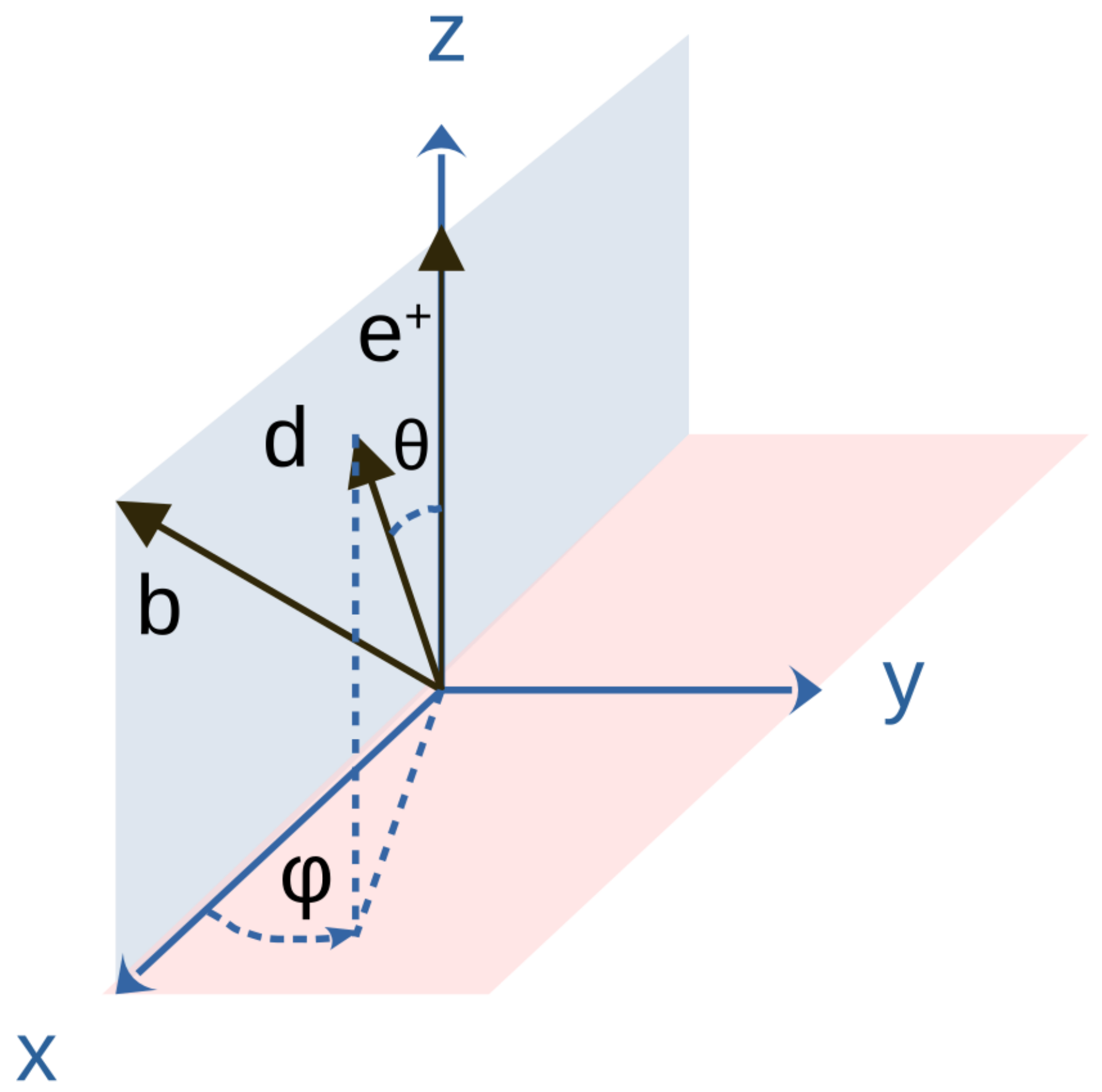}
	\caption{Quantization axis of the top quark spin ($d$-quark momentum direction) in the cluster frame corresponding to the top quark and pseudoscalar mediator decay products.}
	\label{fig:DM_spin_pseudo}
\end{figure}

The difference in the sign in front of the $\sin\theta\cos\phi$ term of the differential cross section for the cases of scalar and pseudoscalar mediators has a clear geometric interpretation. In Figs.~\ref{fig:DM_orientation_scalar}--\ref{fig:DM_spin_pseudo}, the mutual orientation of the production and decay systems of the top quark is shown, as well as the direction of the spin quantization axis ($d$-quark momentum) in the rest frame of the common cluster of the decay products.

\FloatBarrier

\FloatBarrier

In processes with mediator emission, the directions of the final states $b$ and $e^+$ are located closer to each other than in the SM. This is due to the fact that in the common cluster frame the top quark moves and its decay products fly in a narrow sector. The system of these momenta rotates around the direction of the top quark momentum. In turn, the initial system of $u$ and $d$ quarks also rotates and can be oriented differently with respect to the top quark motion axis. The angle $\phi$ is determined by the mutual orientation of the vectors $\mathbf{p}_b$ (from decay) and $\mathbf{p}_d$ (from production), which in the common cluster frame change their kinematics compared to the SM.

In the case of a scalar mediator (Figs.~\ref{fig:DM_orientation_scalar} and \ref{fig:DM_spin_scalar}), the directions of the $d$-quark from production and the $b$-quark from decay are farther from each other. This leads to the predominance of large values of the angle $\phi$ in the differential cross section, and the distribution behaves as ($-\cos\phi$).

For the pseudoscalar mediator (Figs.~\ref{fig:DM_orientation_pseudo} and \ref{fig:DM_spin_pseudo}), the vector $d$ (quantization axis) is oriented closer to the direction of the final state $b$, and small values of $\phi$ predominate, while the distribution is proportional to ($+\cos\phi$).

\FloatBarrier

\FloatBarrier

\section{Kinematic variables for determining the mediator type}
\label{sec:parton_level}

At the first stage, all calculations were performed at the parton level using the CompHEP program~\cite{Boos:2004kh}. This allowed us to identify the key features of the kinematic distributions, free from the influence of hadronization processes and detector response effects. To confirm the efficiency of the method over a wide range of mediator masses, we performed Monte Carlo simulation for two key values. 

\begin{figure}[H]
	\centering
	\includegraphics[width=0.48\textwidth]{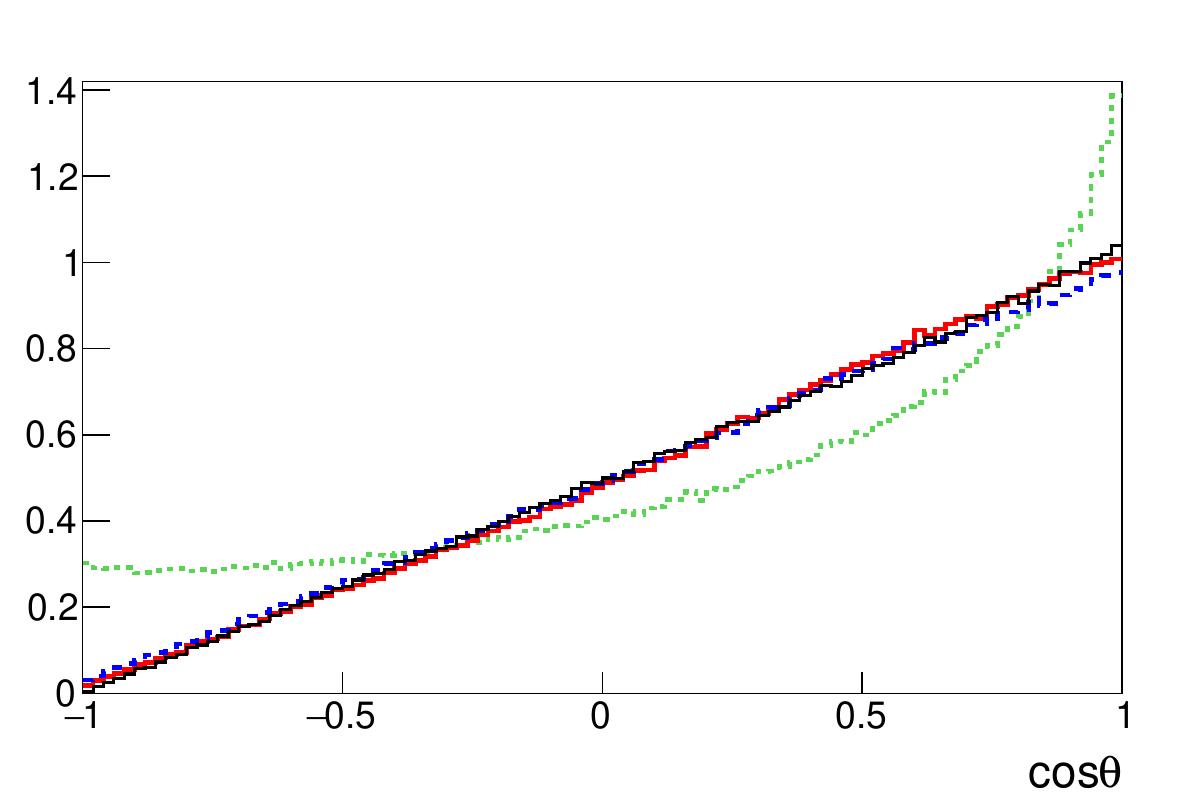}
	\caption{Normalized parton-level distribution in $\cos\theta$ in the common cluster frame of the $t$-quark and mediator decay products: scalar (red solid line), pseudoscalar (blue dashed line), vector (green dotted line). The SM case is shown by the black solid line. Mediator mass 400~GeV.}
	\label{fig:DM_partonic_cos_1}
\end{figure}
\FloatBarrier

\begin{figure}[H]
	\centering
	\includegraphics[width=0.48\textwidth]{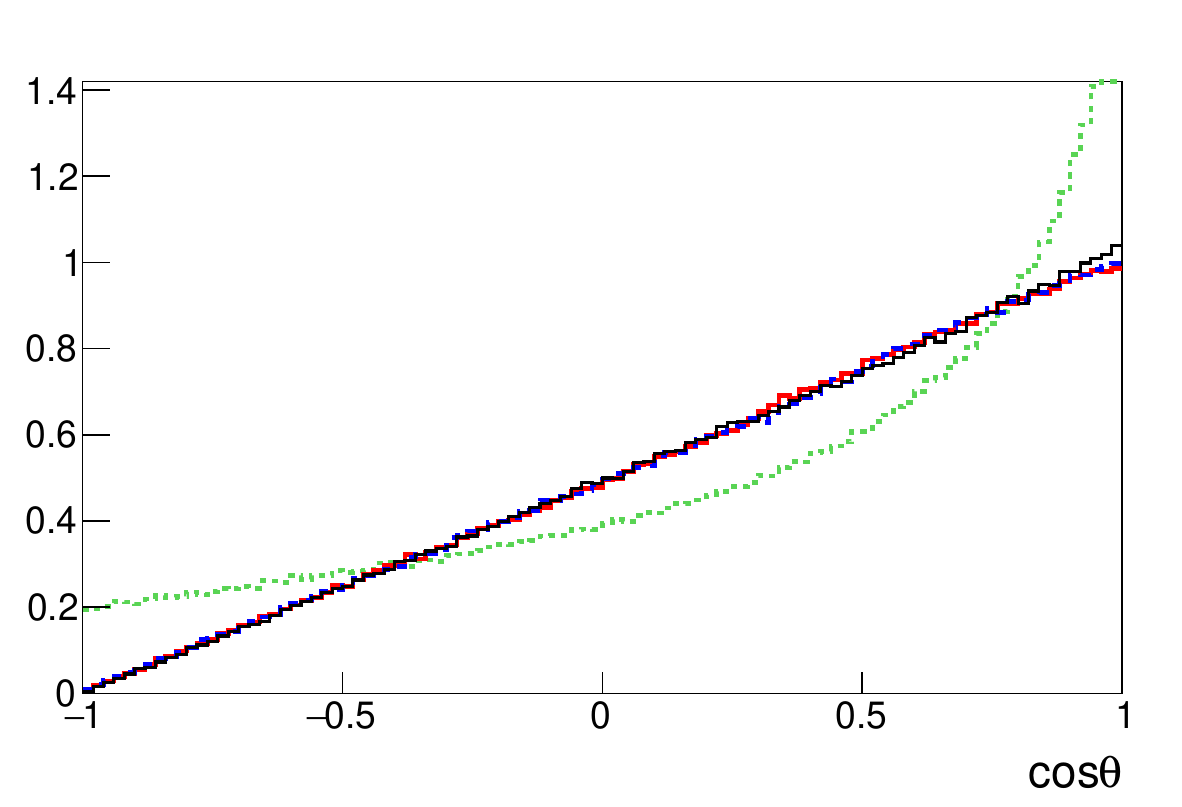}
	\caption{Normalized parton-level distribution in $\cos\theta$ in the common cluster frame of the $t$-quark and mediator decay products: scalar (red solid line), pseudoscalar (blue dashed line), vector (green dotted line). The SM case is shown by the black solid line. Mediator mass 1500~GeV.}
	\label{fig:DM_partonic_cos_2}
\end{figure}
\FloatBarrier

The first value was chosen as 400~GeV, corresponding to the lower unexcluded bound for scalar and pseudoscalar mediators~\cite{CMS:2024zqs}. The second value was taken three times larger than the previous one and corresponds to the lower unexcluded bound for vector mediators. The corresponding kinematic distributions are shown for all scenarios.

In Figs.~\ref{fig:DM_partonic_cos_1} and~\ref{fig:DM_partonic_cos_2}, the normalized distribution in $\cos\theta$ in the common cluster frame is shown. It can be seen that for the SM, scalar, and pseudoscalar mediators, a direct correlation between the directions of the $d$-quark and positron momenta is preserved, while for the vector mediator this correlation is destroyed. This is explained by the contribution of diagrams with mediator emission from the initial quark lines and the more complex Lorentz structure of the vertex. Thus, the variable $\cos\theta$ allows one to single out only the vector mediator, but does not distinguish between the scalar and pseudoscalar scenarios. For mediator masses of 400~GeV and 1500~GeV, these distributions look identical.

\begin{figure}[H]
	\centering
	\includegraphics[width=0.48\textwidth]{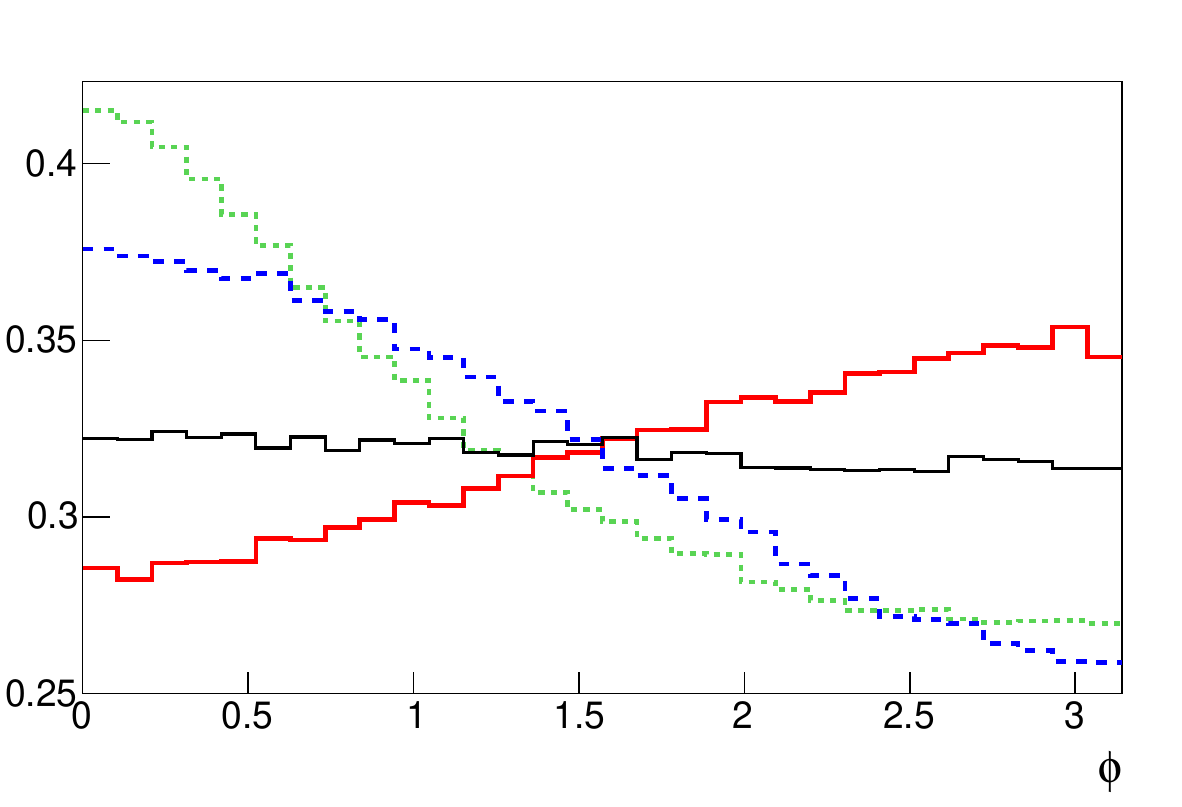}
	\caption{Normalized parton-level distribution in the angle $\phi$ in the common cluster frame of the $t$-quark and mediator decay products: scalar (red solid line), pseudoscalar (blue dashed line), vector (green dotted line). The SM case is shown by the black solid line. Mediator mass 400~GeV.}
	\label{fig:DM_partonic_phi_1}
\end{figure}
\FloatBarrier

\begin{figure}[H]
	\centering
	\includegraphics[width=0.48\textwidth]{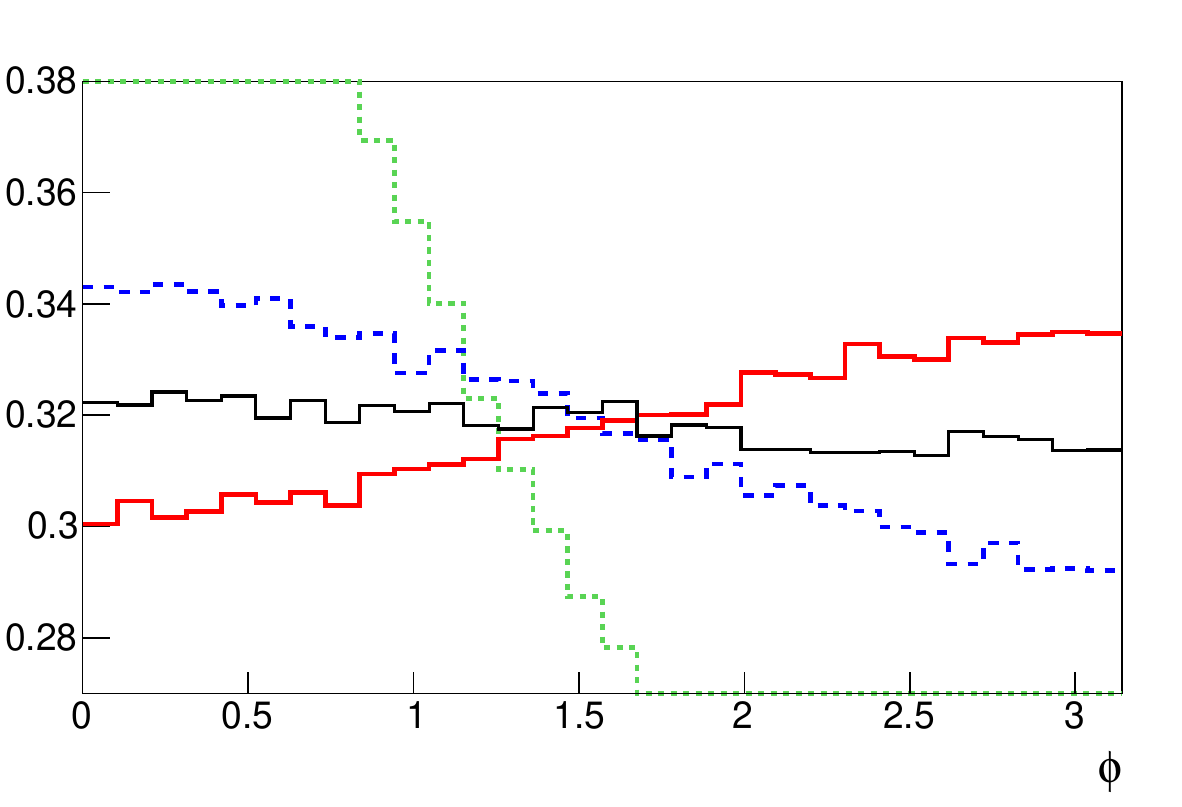}
	\caption{Normalized parton-level distribution in the angle $\phi$ in the common cluster frame of the $t$-quark and mediator decay products: scalar (red solid line), pseudoscalar (blue dashed line), vector (green dotted line). The SM case is shown by the black solid line. Mediator mass 1500~GeV.}
	\label{fig:DM_partonic_phi_2}
\end{figure}
\FloatBarrier

\begin{figure}[H]
	\centering
	\includegraphics[width=0.48\textwidth]{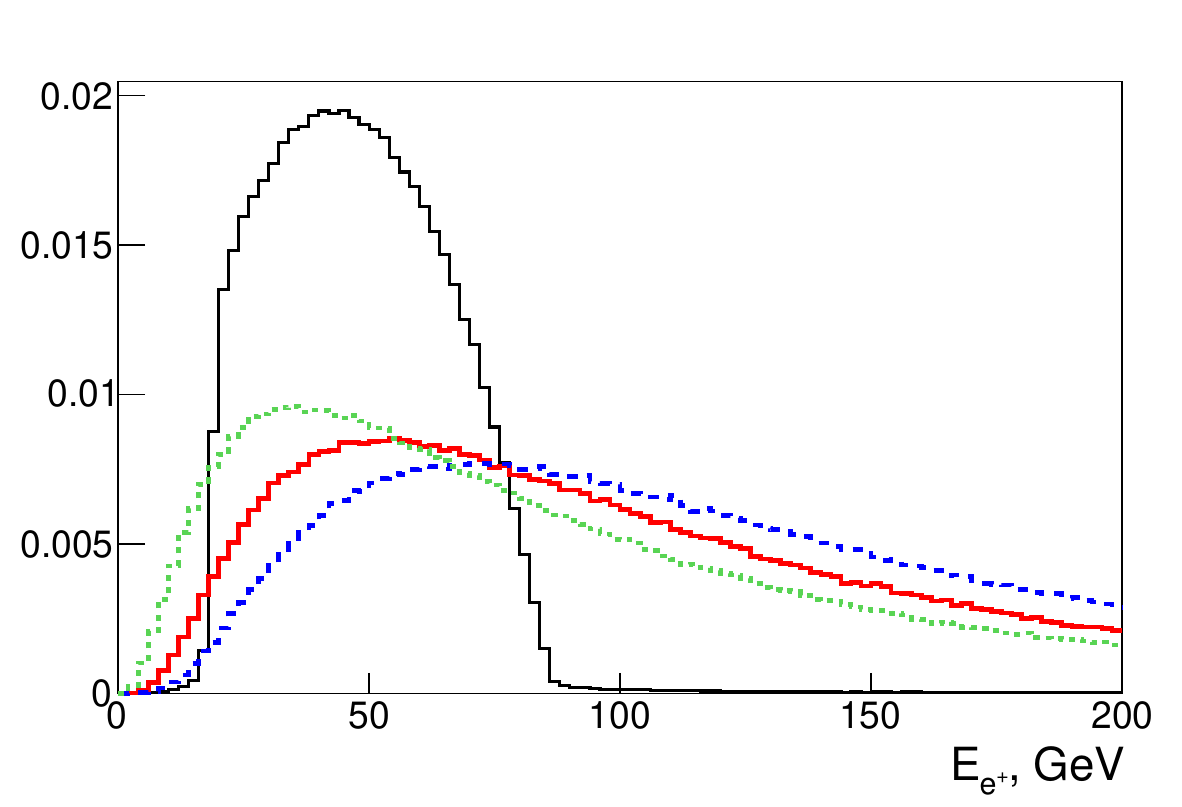}
	\caption{Normalized  parton-level positron energy distribution in the common cluster frame of the $t$-quark and mediator decay products: scalar (red solid line), pseudoscalar (blue dashed line), vector (green dotted line). The SM case is shown by the black solid line. Mediator mass 400~GeV.}
	\label{fig:DM_partonic_energy_1}
\end{figure}
\FloatBarrier

\begin{figure}[H]
	\centering
	\includegraphics[width=0.48\textwidth]{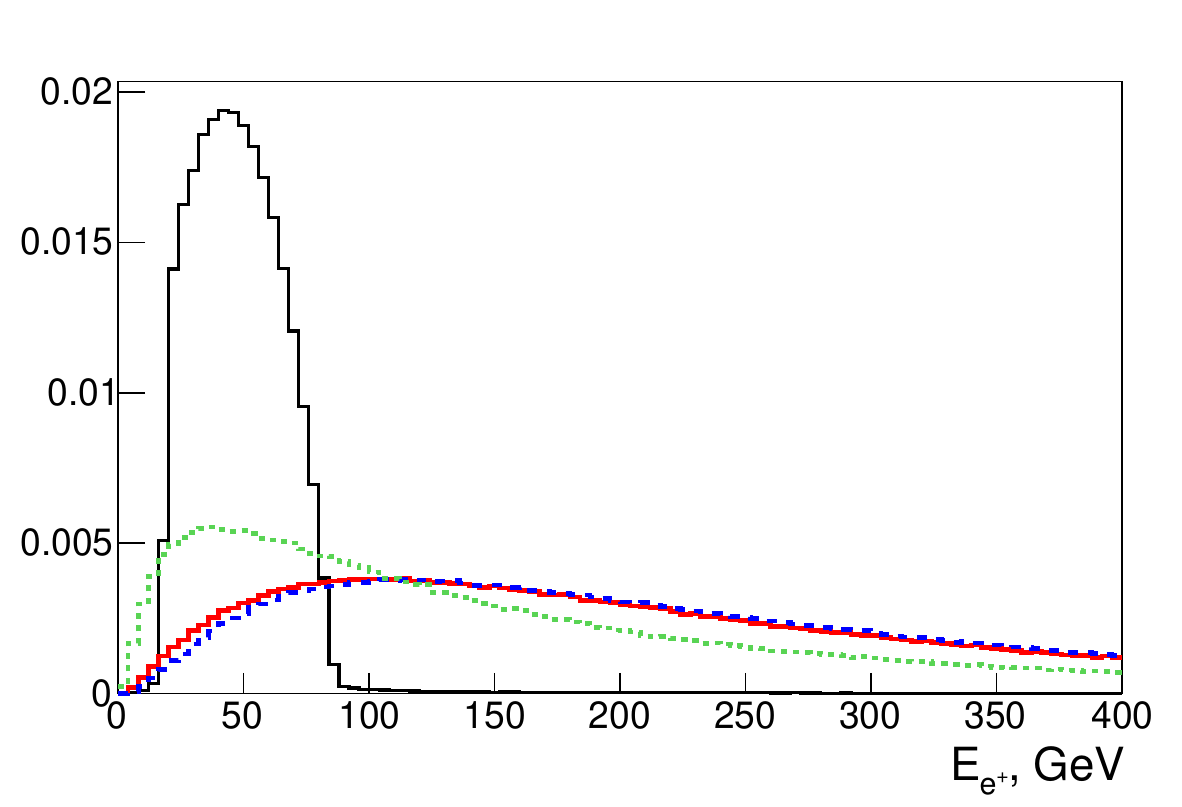}
	\caption{Normalized  parton-level positron energy distribution in the common cluster frame of the $t$-quark and mediator decay products: scalar (red solid line), pseudoscalar (blue dashed line), vector (green dotted line). The SM case is shown by the black solid line. Mediator mass 1500~GeV.}
	\label{fig:DM_partonic_energy_2}
\end{figure}
\FloatBarrier

The distribution in the angle $\phi$ (Figs.~\ref{fig:DM_partonic_phi_1} and~\ref{fig:DM_partonic_phi_2}) is the most informative. In the SM it is constant, for the scalar mediator it has a ($-\cos\phi$) profile, for the pseudoscalar one ($+\cos\phi$), and for the vector one it demonstrates a more complex behavior.

The appearance of the dependence on $\phi$ is due to the fact that in the common cluster frame the top quark has a nonzero momentum, which leads to a change in the kinematics of the $b$-quark and positron. The angle $\phi$ is determined by the mutual orientation of the $b$ and $e^+$ momentum directions relative to the $d$-quark. The change in these directions leads to a nonmonotonic dependence of the differential cross section on the angle $\phi$. The differences in the directions of $b$ and $e^+$ relative to $d$ for mediators with different spin-parity allow one to unambiguously identify the mediator type. Upon transition to a mediator mass of 1500~GeV, the distributions for the scalar and pseudoscalar mediators are slightly smoothed, but on the whole retain their profile, while the distribution for the vector mediator changes its behavior drastically and shifts towards small values of the angle $\phi$.

The positron energy distribution (Fig.~\ref{fig:DM_partonic_energy_1}) also carries important information. In the SM, the positron energy is bounded by the values $E_{\min}=m_W^2/(2m_t)$ and $E_{\max}=m_t/2$, whereas for models with mediators the right tail of the distribution stretches towards larger energy values, and the position of the maximum shifts. These effects are most pronounced for the pseudoscalar mediator. For a mediator mass of 1500~GeV (Fig.~\ref{fig:DM_partonic_energy_2}), the shift of the distribution maxima relative to the SM case is even more enhanced.

Thus, the most serious changes upon transition to larger masses occurred for the distributions in the angle $\phi$ in the case of the vector mediator. At the same time, the distributions corresponding to the scalar and pseudoscalar mediators retained their character, which is consistent with the obtained analytical expressions.

\section{Mediator mass reconstruction}
\label{sec:mass_var_parton}

The separation of neutrino and mediator momenta is difficult to implement in practice, since the number of kinematic equations that can be written is less than the number of unknown momentum components of the particles. In this case, the problem of measuring the mediator mass also seems unsolvable. However, if one uses the kinematic features of this process, it is possible to partially reconstruct the missing information.

\begin{figure}[H]
	\centering
	\includegraphics[width=0.48\textwidth]{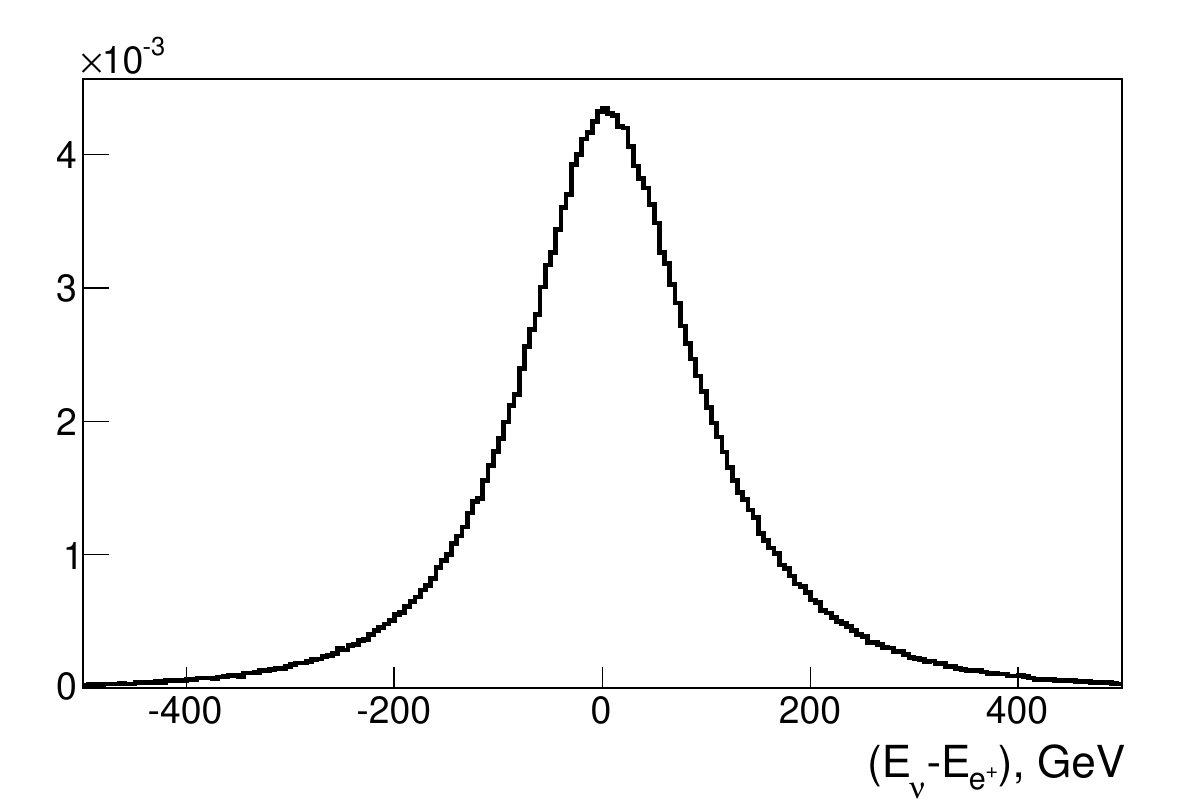}
	\caption{Distribution of the difference between the neutrino and positron energies in the rest frame of the cluster corresponding to the top quark and scalar mediator decay products.}
	\label{fig:DM_nu_e_diff_1}
\end{figure}

\begin{figure}[H]
	\centering
	\includegraphics[width=0.48\textwidth]{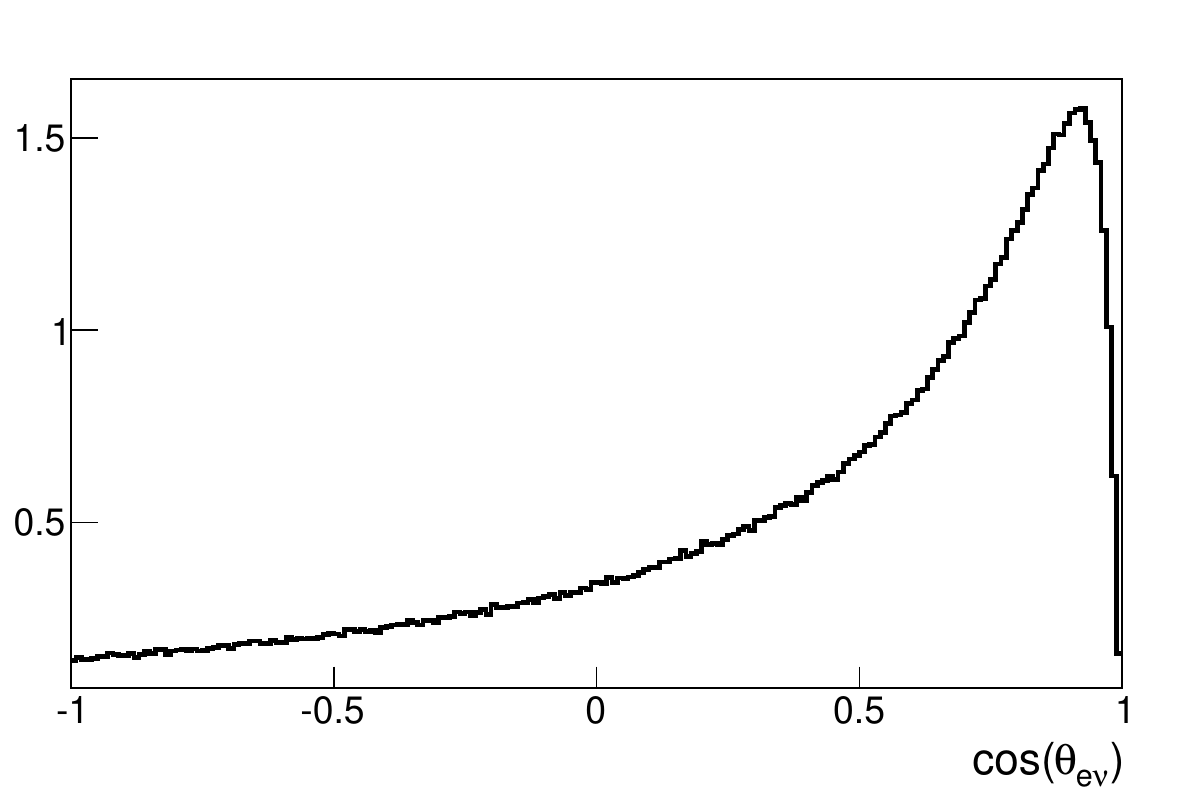}
	\caption{Distribution in the cosine of the angle between the neutrino and positron momentum directions in the rest frame of the cluster corresponding to the top quark and scalar mediator decay products.}
	\label{fig:DM_nu_e_diff_2}
\end{figure}

In the rest frame of the cluster corresponding to the top quark and mediator decay products, the top quark and mediator fly in opposite directions, as shown in Figs.~\ref{fig:DM_orientation_scalar} and~\ref{fig:DM_spin_pseudo}. The moduli of their three-dimensional momenta are equal to each other. In this frame, in the case of a sufficiently large mediator mass, the top quark has a large momentum, part of which it transfers during decay to the $W$-boson, which in turn transfers it to the positron and neutrino. Thus, in this frame, the positron and neutrino fly in the same direction in a narrow sector, and their energy values are close to each other. The distribution for the difference of the positron and neutrino energies (Fig.~\ref{fig:DM_nu_e_diff_1}), as well as the distribution for the cosine of the angle between their momenta (Fig.~\ref{fig:DM_nu_e_diff_2}), confirm this assumption.

\FloatBarrier

\FloatBarrier

Thus, in most cases the neutrino and positron momenta coincide, and we can replace the neutrino momentum by the positron momentum in this frame. This approximation is applicable for $m_\phi \gg m_t$ and is not exact. According to the relativistic relation, the mediator mass is equal to:
\begin{equation}
	M_{\phi} = \sqrt{E_{\phi}^2 - \mathbf{p}_{\phi}^2} = \sqrt{(E_{\text{miss}} - E_{\nu})^2 - (\mathbf{p}_{\text{miss}} - \mathbf{p}_{\nu})^2}.
\end{equation}
Replacing the neutrino four-momentum by the positron four-momentum, we obtain:
\begin{equation}\label{eq:Mphi1}
	M_{\phi} = \sqrt{(E_{\text{miss}} - E_{e^+})^2 - (\mathbf{p}_{\text{miss}} - \mathbf{p}_{e^+})^2}.
\end{equation}
Since the three-dimensional momentum of the mediator is equal to the total momentum of the top quark decay products with the opposite sign, and the neutrino momentum is equal to the positron momentum, we can write:
\begin{equation}\label{eq:Mphi2}
	M_{\phi} = \sqrt{(E_{\text{miss}} - E_{e^+})^2 - (\mathbf{p}_{b} + 2\mathbf{p}_{e^+})^2},
\end{equation}
where $E_{\text{miss}}$ is the total missing energy in the rest frame of the cluster corresponding to the top quark and mediator decay products.

Using the obtained expression, one can determine the reconstructed mediator mass for the majority of the simulated Monte Carlo events. The position of the maximum of this distribution will be the exact value of the mediator mass. The scheme works when the mediator is sufficiently heavy and in the cluster frame ($W$-boson, $b$-quark, mediator) the $W$-boson flies fast.

\begin{figure}[H]
	\centering
	\includegraphics[width=0.48\textwidth]{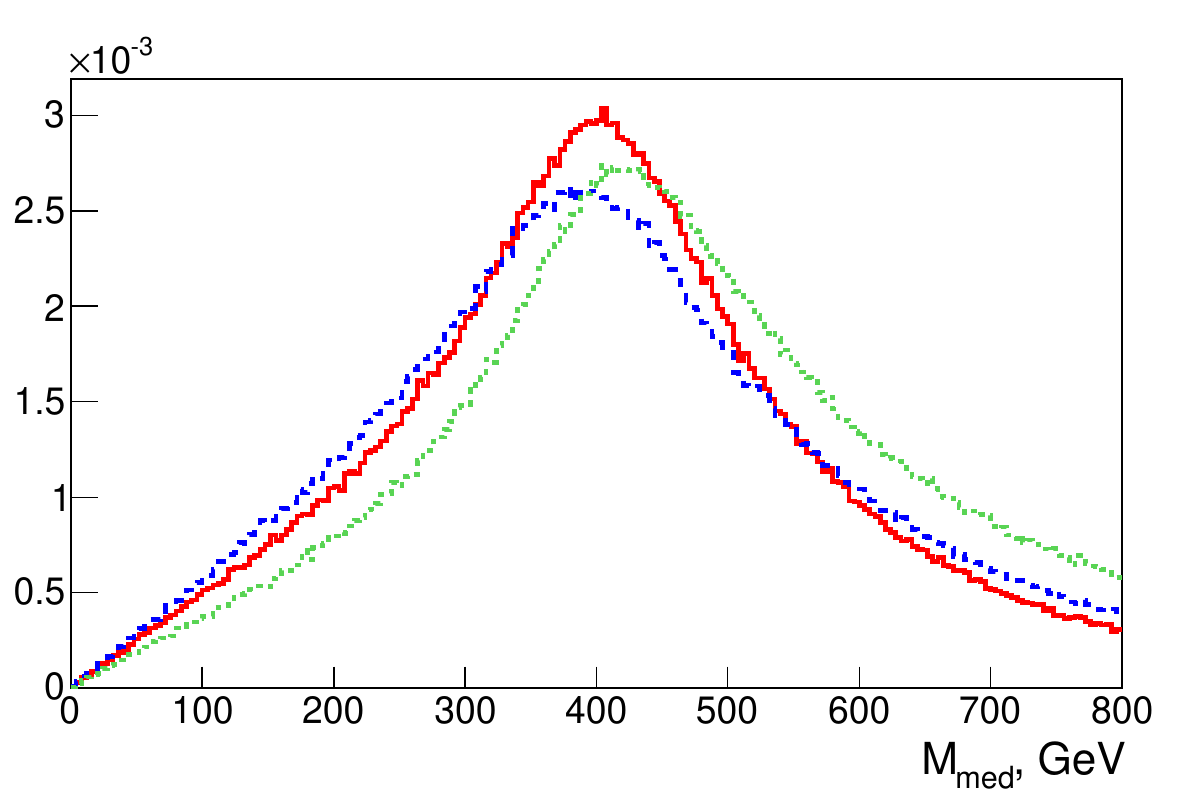}
	\caption{Reconstructed mediator mass in the rest frame of the cluster corresponding to the top quark and mediator decay products: scalar (red solid line), pseudoscalar (blue dashed line), vector (green dotted line). Mediator mass 400~GeV.}
	\label{fig:DM_mass_reco_1}
\end{figure}
\FloatBarrier

\begin{figure}[H]
	\centering
	\includegraphics[width=0.48\textwidth]{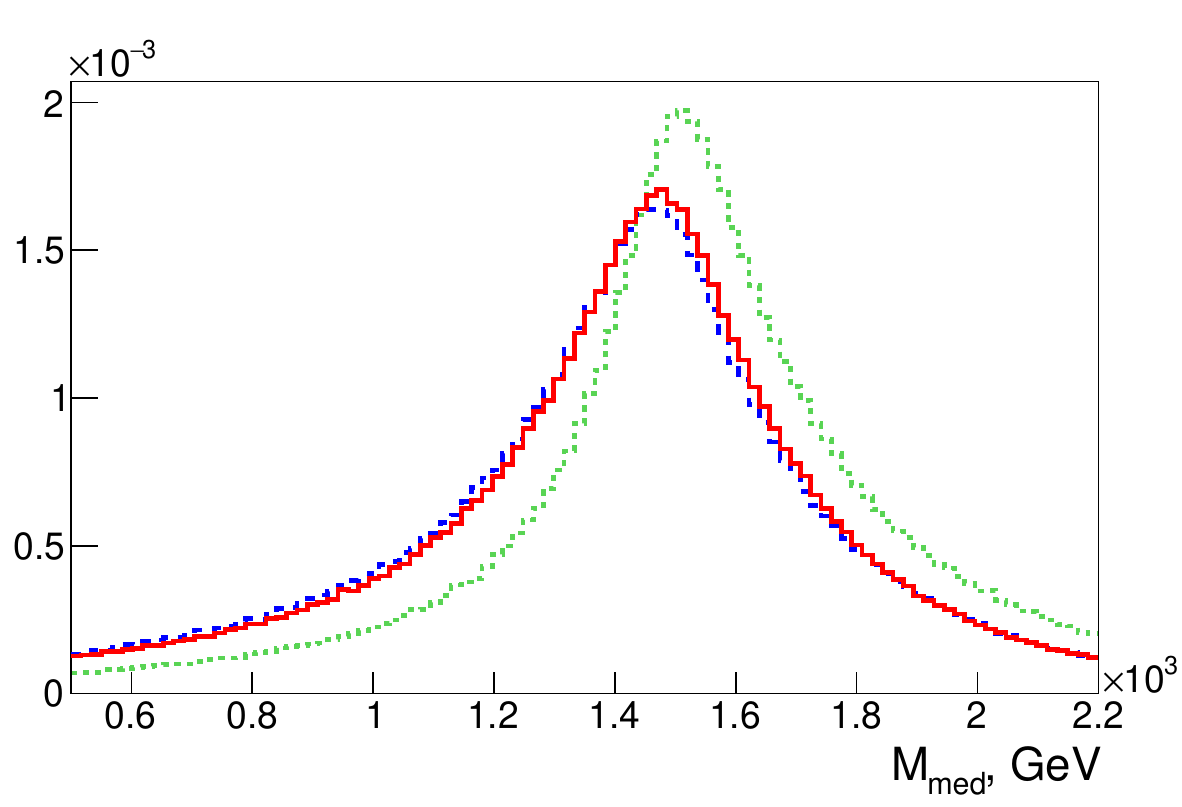}
	\caption{Reconstructed mediator mass in the rest frame of the cluster corresponding to the top quark and mediator decay products: scalar (red solid line), pseudoscalar (blue dashed line), vector (green dotted line). Mediator mass 1500~GeV.}
	\label{fig:DM_mass_reco_2}
\end{figure}
\FloatBarrier

In Fig.~\ref{fig:DM_mass_reco_1} (red solid line), the distribution of the reconstructed mass of the scalar mediator is shown. In the same figure (blue dashed line), the distribution of the reconstructed mass of the pseudoscalar mediator is shown. It can be seen that the positions of the distribution peaks coincide with the input mediator mass of 400~GeV. Although the method was developed for processes involving scalar mediators, it can also be tested in the case of a vector mediator. In Fig.~\ref{fig:DM_mass_reco_1} (green dotted line), the distribution of the reconstructed mass of the vector mediator is shown. It can be seen that the distribution peak has shifted somewhat towards larger values. This is due to the fact that in the process involving the vector mediator, diagrams with mediator emission from the initial $u$- and $b$-quark lines, as well as from the final $d$-quarks, contribute, thereby violating the kinematics of the main subprocess. The corresponding distributions for a mediator mass of 1500~GeV (Fig.~\ref{fig:DM_mass_reco_2}) show that for the scalar and pseudoscalar mediators the reconstruction method continues to work accurately, while the peak for the vector mediator has shifted even more strongly towards larger values.

\begin{figure}[H]
	\centering
	\includegraphics[width=0.48\textwidth]{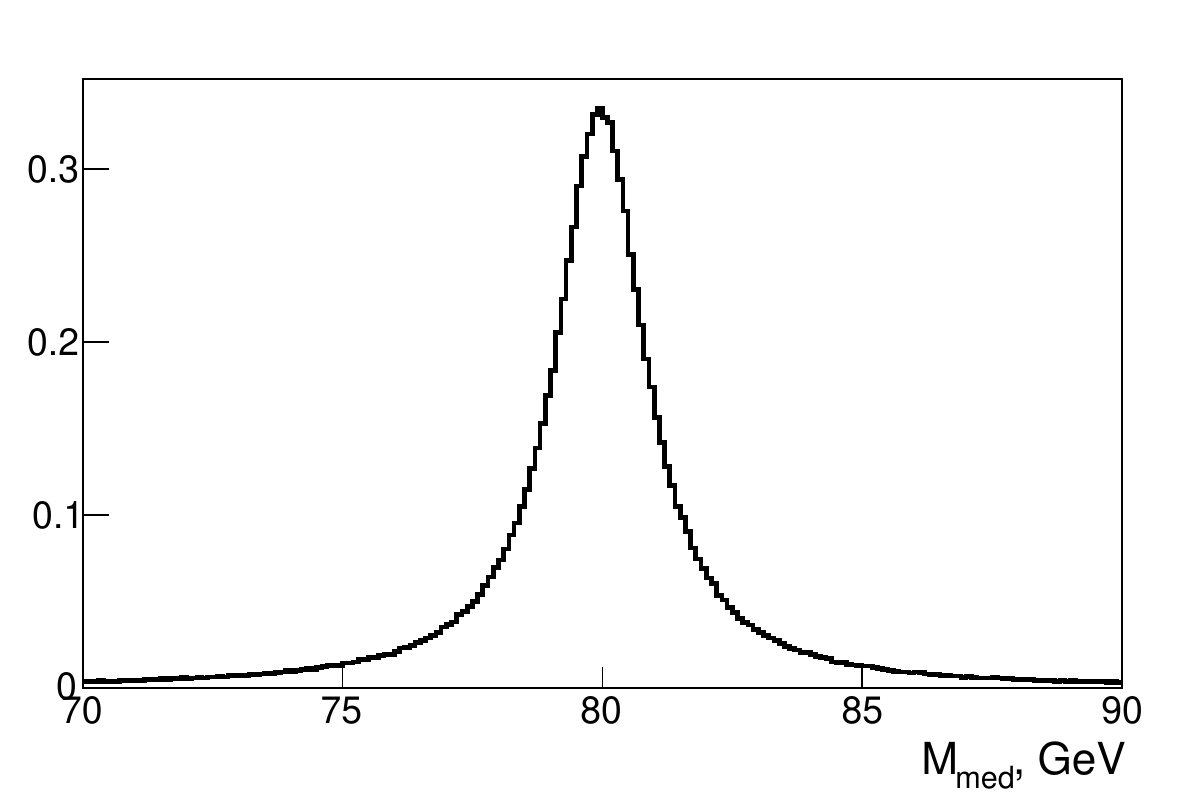}
	\caption{Result of the mediator mass reconstruction algorithm applied to Monte Carlo samples corresponding to the Standard Model.}
	\label{fig:DM_mass_SM}
\end{figure}

If the method is applied to SM samples, a distribution with a maximum at the $W$-boson mass value is obtained (Fig.~\ref{fig:DM_mass_SM}). Such an application makes no sense, as it goes beyond the applicability of the method. Before using the method, a preliminary selection of events with large missing energy must be performed.

\FloatBarrier

In addition to measuring the mediator mass, the described method can be used to select events involving the mediator and dark matter. If one imposes a kinematic cut on the variable (\ref{eq:Mphi2}) at a value equal to the top quark mass and discards all events corresponding to smaller values, one can effectively separate events with new physics from SM events.

\section{Simulation taking into account the detector response}
\label{sec:full_simulation}

To verify the feasibility of the proposed method under real experimental conditions, a full simulation was performed, including the account of hadronization, parton showers, and the response of the CMS detectors at the LHC. For parton-level matrix element calculations and event generation, we used CompHEP 4.5.2 with CTEQ6L PDF sets. The events were then passed to PYTHIA 8.245~\cite{Sjostrand:2007gs} for parton shower and hadronization simulation, and further to Delphes 3.4.2~\cite{deFavereau:2013fsa} for detector response simulation. The kinematic distributions were constructed in the ROOT analysis package~\cite{Brun:1997pa}. 

\begin{figure}[H]
	\centering
	\includegraphics[width=0.48\textwidth]{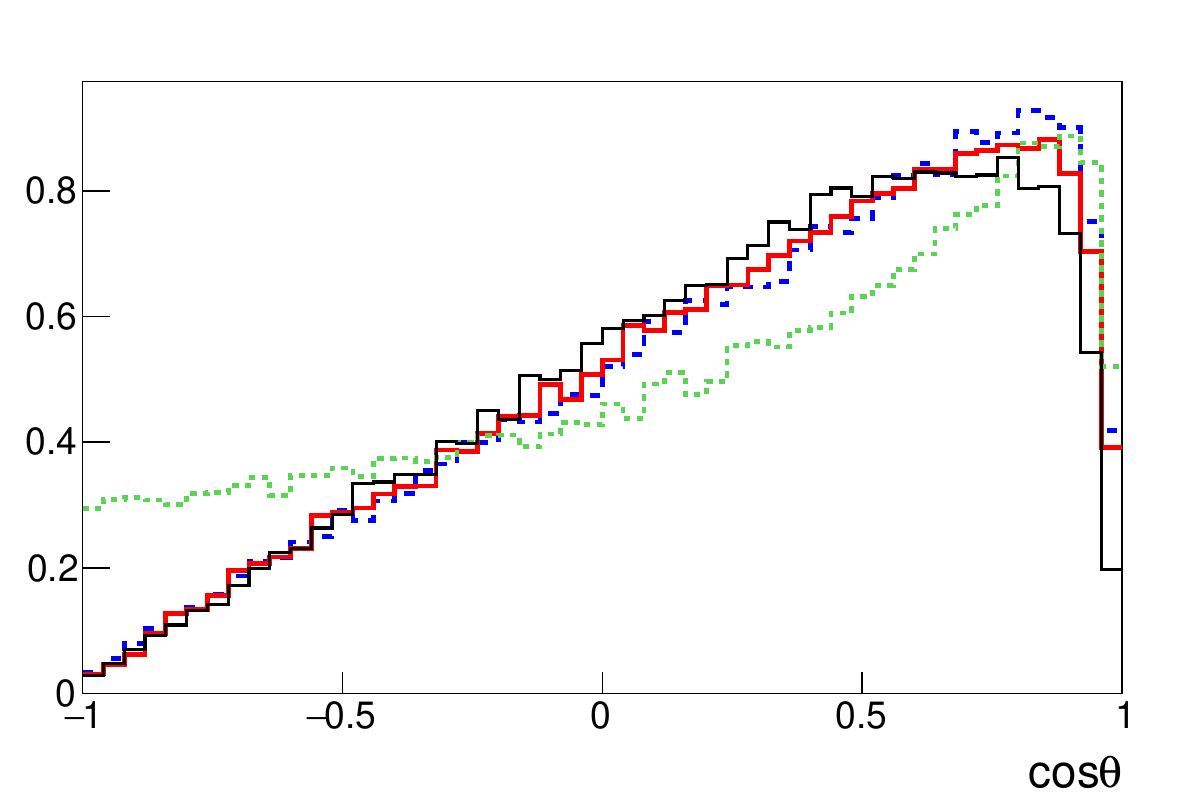}
	\caption{Normalized $\cos\theta$ distribution after full simulation in the common cluster frame of the $t$-quark and mediator decay products: scalar (red solid line), pseudoscalar (blue dashed line), vector (green dotted line). The SM case is shown by the black solid line. Mediator mass 400~GeV.}
	\label{fig:DM_full_cos_400}
\end{figure}
\FloatBarrier

\begin{figure}[H]
	\centering
	\includegraphics[width=0.48\textwidth]{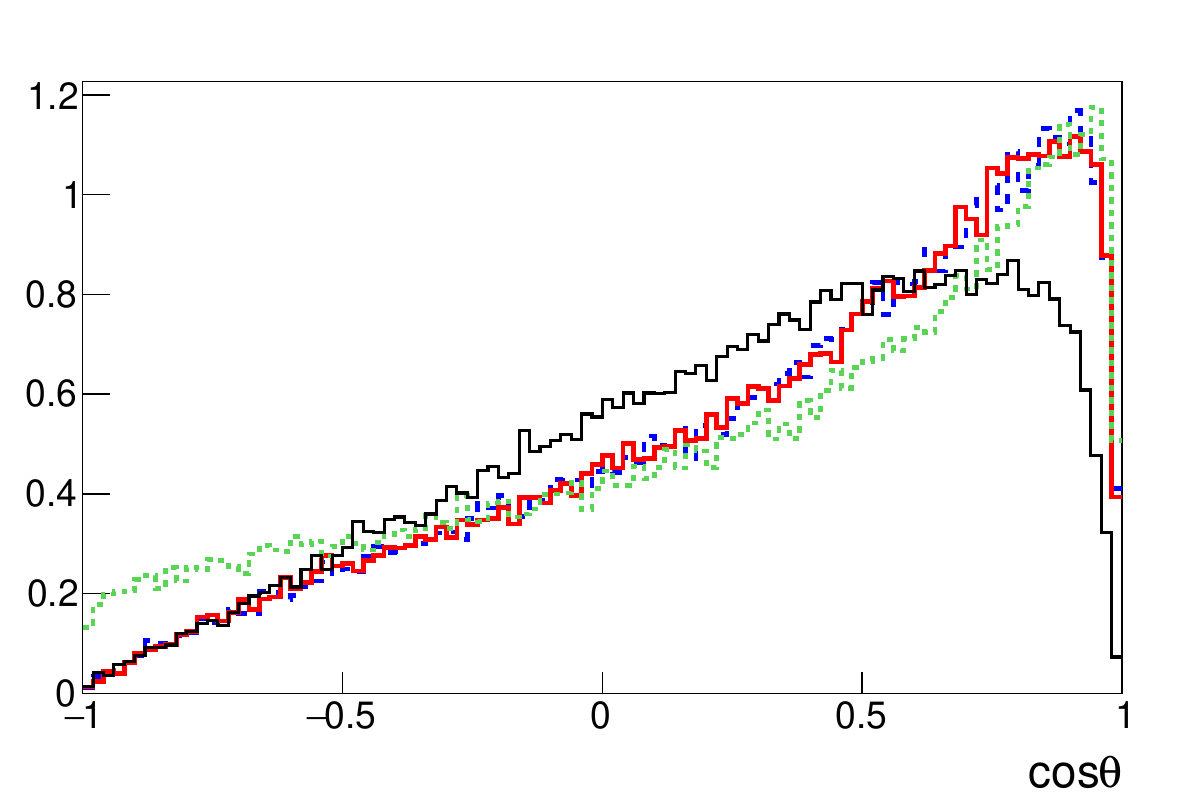}
	\caption{Normalized $\cos\theta$ distribution after full simulation in the common cluster frame of the $t$-quark and mediator decay products: scalar (red solid line), pseudoscalar (blue dashed line), vector (green dotted line). The SM case is shown by the black solid line. Mediator mass 1500~GeV.}
	\label{fig:DM_full_cos_1500}
\end{figure}
\FloatBarrier

Our task was to conduct a comparative study at the most general qualitative level without fixing a specific model and values of the mediator interaction parameters. Therefore, for each scenario we generated 50k events and constructed normalized kinematic distributions. We took a relatively small number of events because we assume that only selected events with large missing energy are studied, and therefore the SM contributions, as well as interference with the SM, can be neglected. In each scenario, two cases were considered: with a mediator mass of 400~GeV (lower unexcluded bound for scalar and pseudoscalar mediators) and 1500~GeV (lower unexcluded bound for vector mediators). The simulation results with distributions in the $\cos\theta$, angle $\phi$, positron energy, and the variable $M_{\phi}$ are presented in Figs.~\ref{fig:DM_full_energy_400}--\ref{fig:DM_full_mass_1500}.

\begin{figure}[H]
	\centering
	\includegraphics[width=0.48\textwidth]{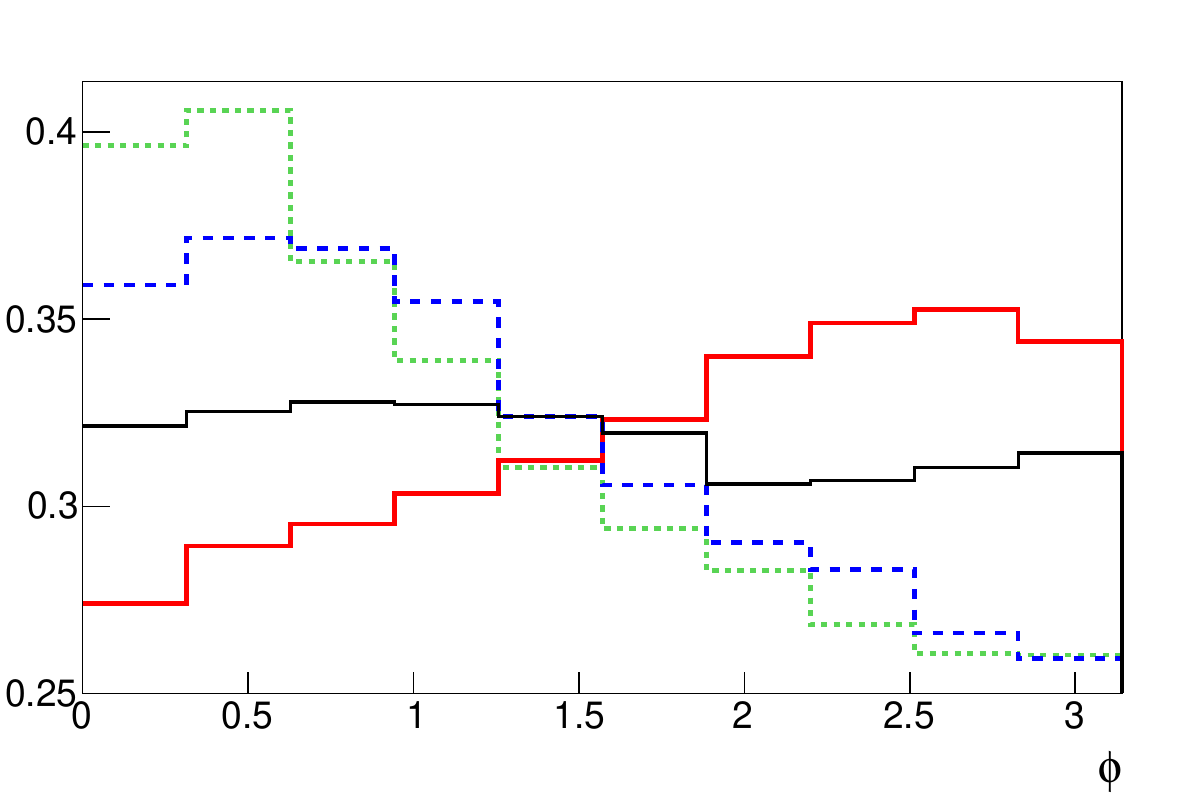}
	\caption{Normalized $\phi$ distribution after full simulation in the common cluster frame of the $t$-quark and mediator decay products: scalar (red solid line), pseudoscalar (blue dashed line), vector (green dotted line). The SM case is shown by the black solid line. Mediator mass 400~GeV.}
	\label{fig:DM_full_phi_400}
\end{figure}
\FloatBarrier

\begin{figure}[H]
	\centering
	\includegraphics[width=0.48\textwidth]{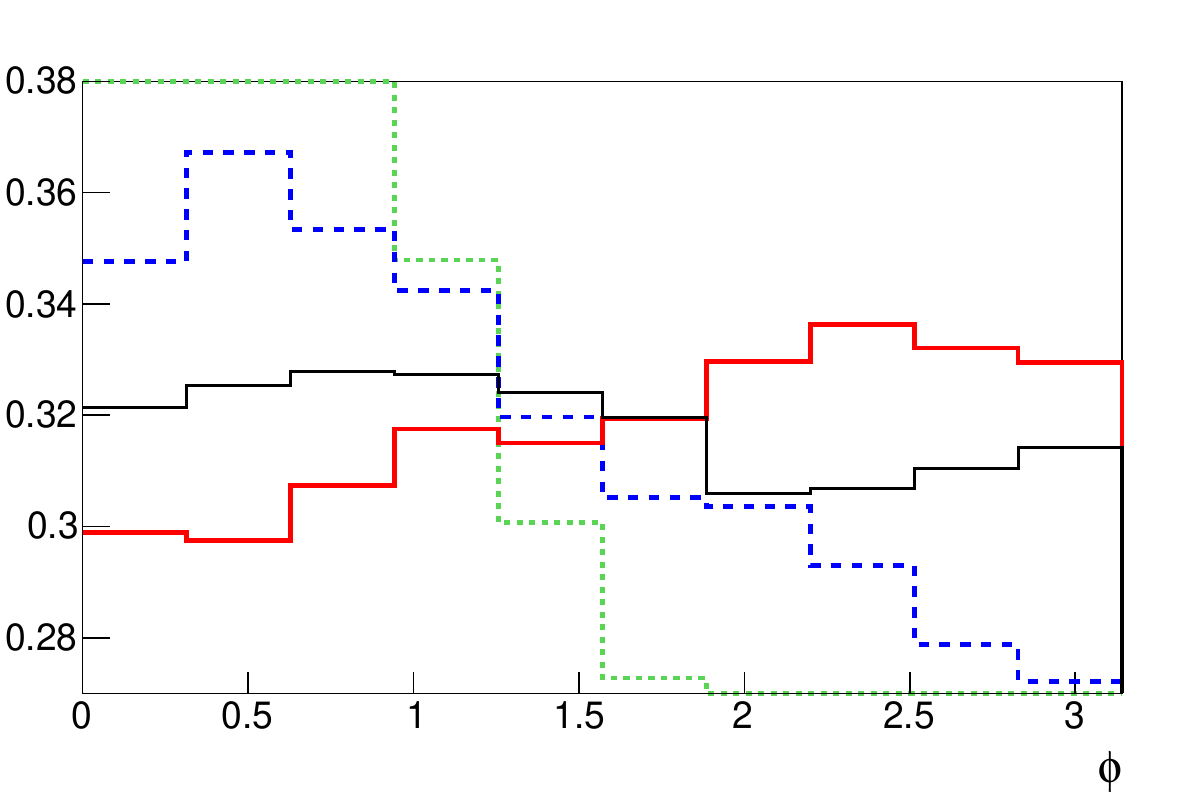}
	\caption{Normalized $\phi$ distribution after full simulation in the common cluster frame of the $t$-quark and mediator decay products: scalar (red solid line), pseudoscalar (blue dashed line), vector (green dotted line). The SM case is shown by the black solid line. Mediator mass 1500~GeV.}
	\label{fig:DM_full_phi_1500}
\end{figure}
\FloatBarrier

\begin{figure}[H]
	\centering
	\includegraphics[width=0.48\textwidth]{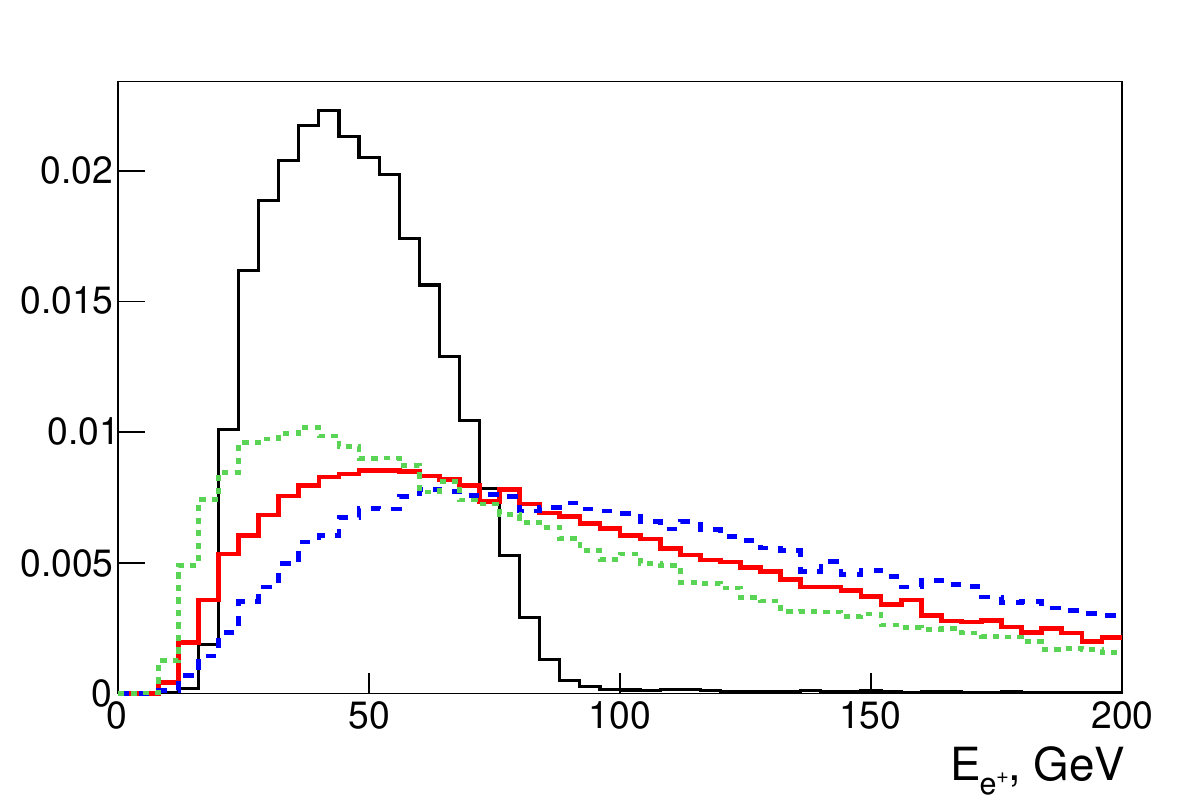}
	\caption{Normalized positron energy distribution after full simulation in the common cluster frame of the $t$-quark and mediator decay products: scalar (red solid line), pseudoscalar (blue dashed line), vector (green dotted line). The SM case is shown by the black solid line. Mediator mass 400~GeV.}
	\label{fig:DM_full_energy_400}
\end{figure}
\FloatBarrier

\begin{figure}[H]
	\centering
	\includegraphics[width=0.48\textwidth]{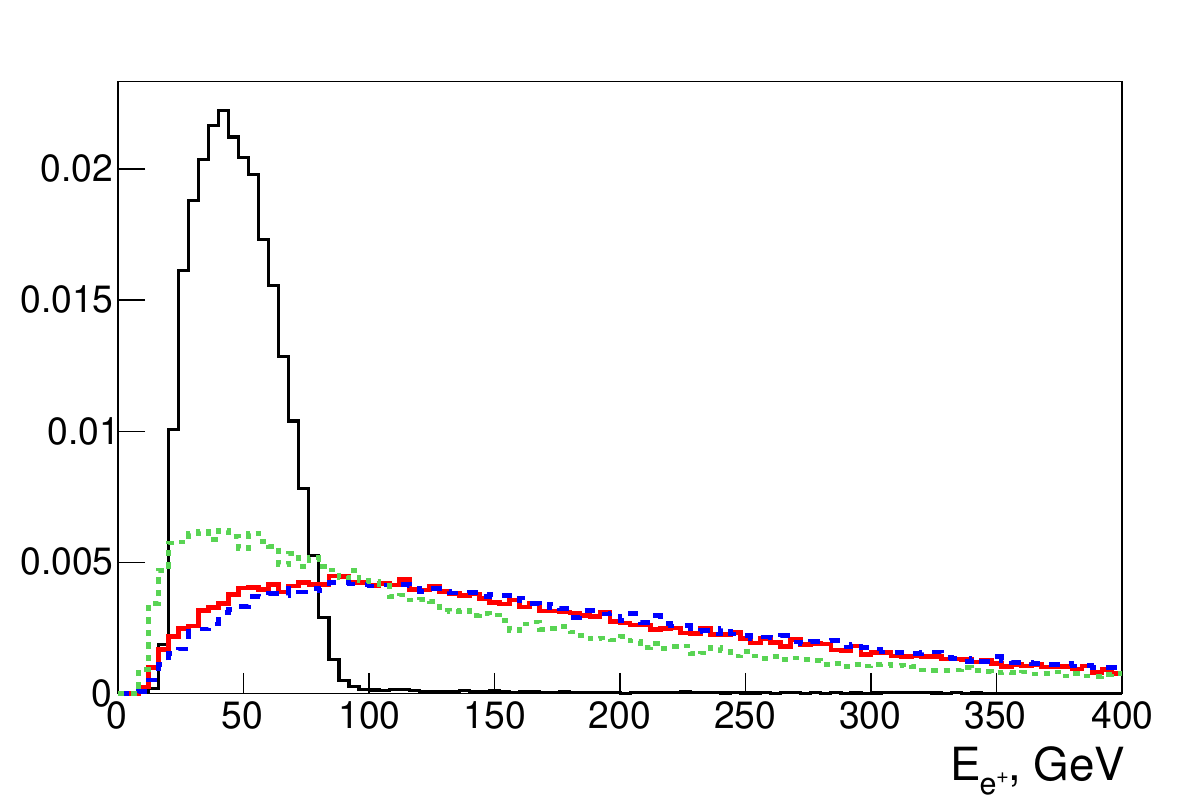}
	\caption{Normalized positron energy distribution after full simulation in the common cluster frame of the $t$-quark and mediator decay products: scalar (red solid line), pseudoscalar (blue dashed line), vector (green dotted line). The SM case is shown by the black solid line. Mediator mass 1500~GeV.}
	\label{fig:DM_full_energy_1500}
\end{figure}
\FloatBarrier

The distributions in $\cos\theta$ (Figs.~\ref{fig:DM_full_cos_400} and~\ref{fig:DM_full_cos_1500}) after taking into account detector effects for a mass of 400~GeV become more smoothed in the region closer to 1, but on the whole the correlation behavior ($1+\cos\theta$) for the scalar and pseudoscalar cases is preserved and allows one to distinguish these scenarios from the vector mediator case. However, at a mass of 1500~GeV the differences between scenarios with different mediators significantly weaken.

The distribution in the angle $\phi$ (Figs.~\ref{fig:DM_full_phi_400} and~\ref{fig:DM_full_phi_1500}) after taking into account detector effects becomes somewhat more smeared due to errors in reconstructing the jet directions, but the characteristic behavior ($-\cos\phi$) for the scalar, ($+\cos\phi$) for the pseudoscalar) is preserved, which allows the mediator to be identified. At a mass of 1500~GeV the differences become less pronounced, but are still noticeable.

The positron energy distribution (Figs.~\ref{fig:DM_full_energy_400} and~\ref{fig:DM_full_energy_1500}) reproduces the parton predictions well. For a mass of 400~GeV, the maxima of the distributions for the scalar and pseudoscalar mediators are noticeably shifted to the right relative to the SM, and upon increasing the mass to 1500~GeV this shift becomes even more pronounced, which improves the separation of scenarios.

\begin{figure}[H]
	\centering
	\includegraphics[width=0.48\textwidth]{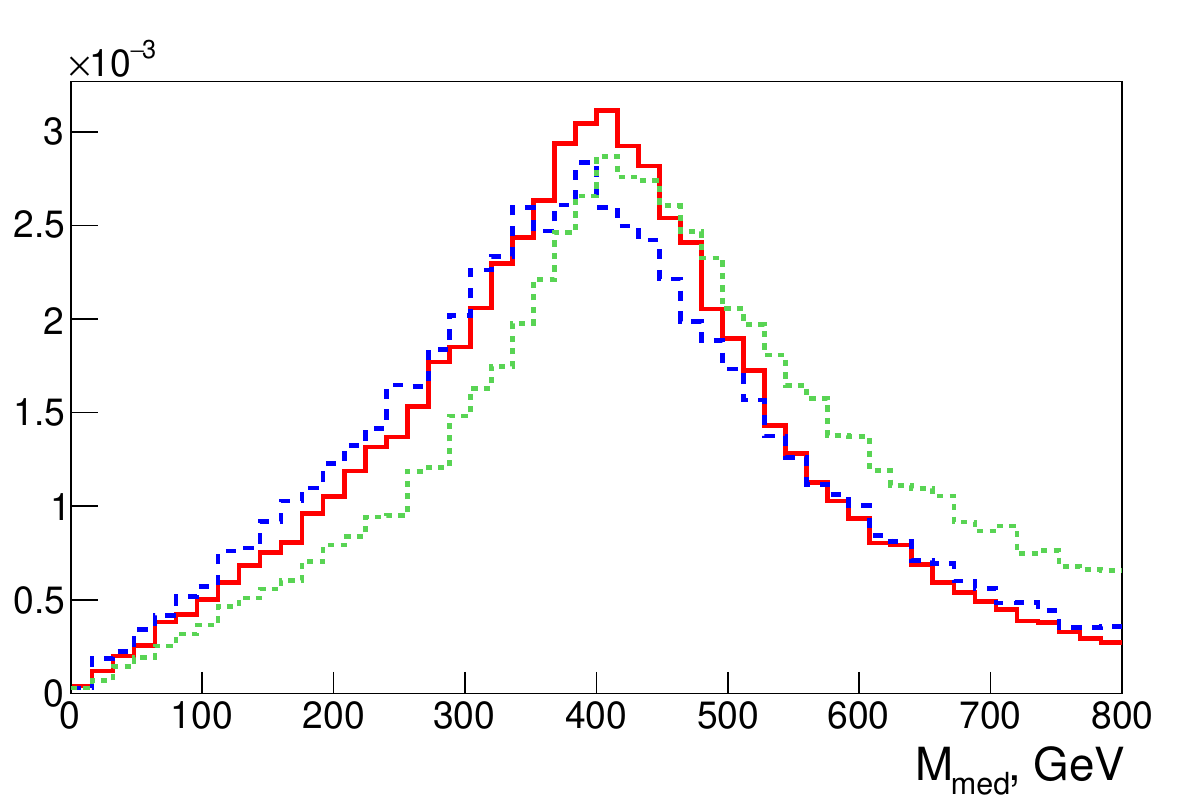}
	\caption{Distribution in the variable $M_{\phi}$ after full simulation in the common cluster frame of the $t$-quark and mediator decay products: scalar (red solid line), pseudoscalar (blue dashed line), vector (green dotted line). The SM case is shown by the black solid line. Mediator mass 400~GeV.}
	\label{fig:DM_full_mass_400}
\end{figure}
\FloatBarrier

\begin{figure}[H]
	\centering
	\includegraphics[width=0.48\textwidth]{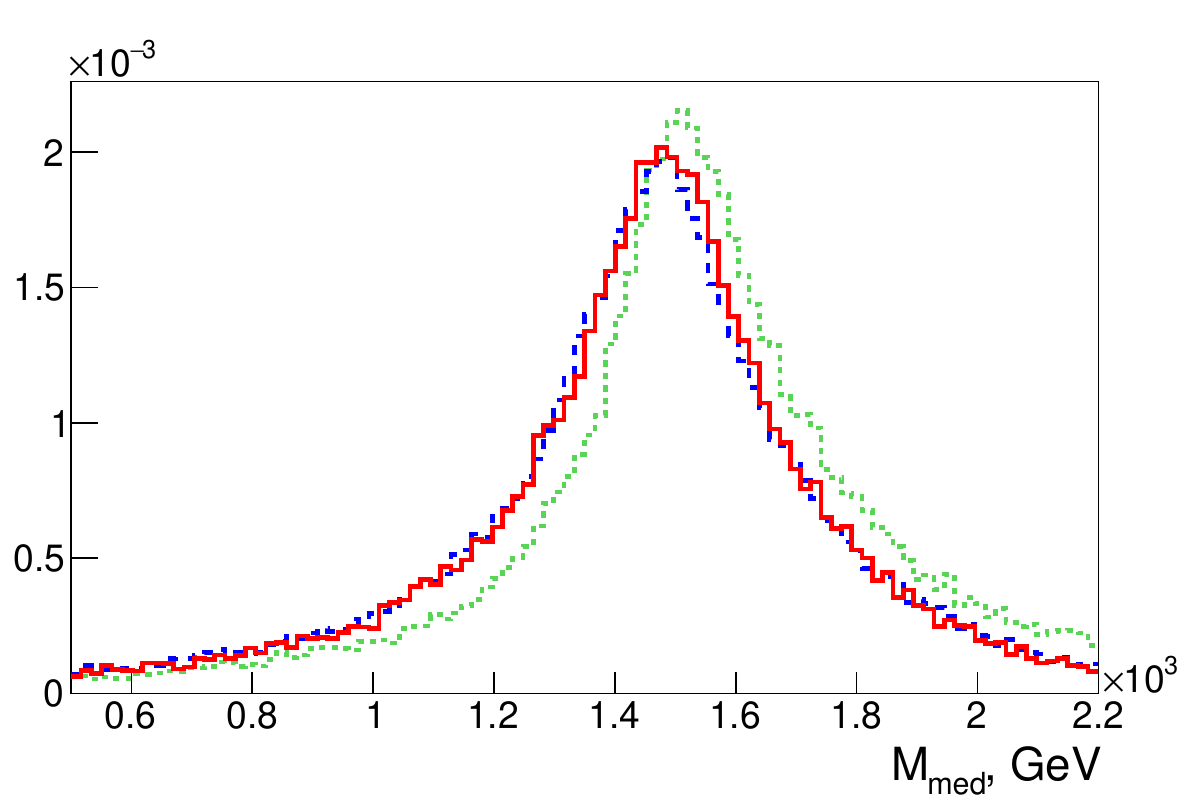}
	\caption{Distribution in the variable $M_{\phi}$ after full simulation in the common cluster frame of the $t$-quark and mediator decay products: scalar (red solid line), pseudoscalar (blue dashed line), vector (green dotted line). The SM case is shown by the black solid line. Mediator mass 1500~GeV.}
	\label{fig:DM_full_mass_1500}
\end{figure}
\FloatBarrier

The distributions in the variable $M_{\phi}$ (Figs.~\ref{fig:DM_full_mass_400} and~\ref{fig:DM_full_mass_1500}) demonstrate clear peaks near the true masses of the scalar and pseudoscalar mediators, while the shift of the peak for the vector mediator to the right is consistent with expectations. Even at a mass of 1500~GeV the peaks remain well distinguishable and allow the mediator mass to be measured with high accuracy.

Thus, the performed simulation has shown that the most effective variables for searching for dark matter mediators are the angle $\phi$, the charged lepton energy, and the variable reconstructing the mass in the rest frame of the common cluster of the top quark and mediator decay products.

\section{Conclusion}

In the present work, differential cross sections describing spin correlations in processes of associated dark matter and single top quark production at the LHC have been obtained. Scenarios with scalar, pseudoscalar, and vector mediators have been investigated in detail. It has been shown that a joint analysis of the distributions in the positron energy, the angle $\phi$, and the variable $M_{\phi}$ in the rest frame of the common cluster of the top quark and mediator decay products allows one to effectively identify the mediator type and measure its mass. The developed mass reconstruction method, using the approximation of equal positron and neutrino momenta, has demonstrated high accuracy both at the parton level and after full simulation taking into account the detector response. The proposed approach can be directly applied in LHC experiments for the search and study of dark matter particles.

\section{Acknowledgments}

This research was carried out within the framework of the scientific program of the National Center for Physics and Mathematics, project ``Physics of Elementary Particles and Cosmology''.

\bibliographystyle{apsrev4-2}
\bibliography{references}

\end{document}